\documentclass[a4paper,11pt]{article}
\pdfoutput=1 
\usepackage{jheppub} 

\title{Kinetics of Hawking-Page phase transition with the non-Markovian effects}

\author[a,b]{Ran Li,}
\author[b,c,*]{Jin Wang \note[*]{Corresponding author}}
\affiliation[a]{School of Physics, Henan Normal University, Xinxiang 453007, China}
\affiliation[b]{Department of Chemistry, Stony Brook University, Stony Brook, NY 11794, USA}
\affiliation[c]{Department of Physics and Astronomy, Stony Brook University, Stony Brook, NY 11794, USA,}

\emailAdd{liran@htu.edu.cn}
\emailAdd{jin.wang.1@stonybrook.edu}

\abstract{Based on the free energy landscape description of Hawking-Page phase transition, the transition process from the Schwarzschild-anti-de Sitter black hole to the thermal anti-de Sitter space are considered to be stochastic under the thermal fluctuations. If the correlation time of the effective thermal bath is comparable or even longer than the oscillating time of the spacetime state in the potential well on the free energy landscape, the non-Markovian model of the black hole phase transition is required to study the kinetics of the transition processes. The non-Markovian or memory effect is represented by the time dependent friction kernel and the kinetics is then governed by the generalized Langevin equation complemented by the free energy potential. As the concrete examples, we study the effects of the exponentially decay friction kernel and the oscillatory friction kernel on the kinetics of Hawking-Page phase transition. For the exponentially decayed friction, the non-Markovian effects promote the transition process, and for the oscillatory friction, increasing the oscillating frequency also speeds up the transition process. }
\begin{document}

\maketitle

\flushbottom

\section{Introduction}
 
The celebrated conjecture of holographic principle has fascinated the theorists for almost three decades and promoted our understanding of the nature of the quantum gravity \cite{tHooft:1993dmi,Susskind:1994vu}. The AdS/CFT correspondence as a precise realization of holographic principle states that the gravity living in the bulk AdS spacetime can be equivalently described by the conformal field theory on the boundary \cite{Maldacena:1997re,Gubser:1998bc,Witten:1998qj}. Studying the thermodynamics of AdS black holes from the viewpoint of statistical physics appears to be essential to our understanding of the AdS/CFT correspondence. In particular, the Hawking-Page transition \cite{Hawking:1982dh}, which is a first order phase transition between the thermal AdS space and the large Schwarzschild-AdS (SAdS) black hole, has been properly explained as the confinement/deconfinement transition in quantum chromodynamics (QCD) \cite{Witten:1998zw}.

The Hawking-Page phase transition has been widely studied in the literature from holography to modified gravity theories \cite{Cappiello:2001tf,Cho:2002hq,Birmingham:2002ph,Biswas:2003sn,Cai:2007zw,Cai:2007wz,Cai:2007vv,Nicolini:2011dp,Eune:2013qs,Spallucci:2013jja,Zhang:2015wna,Yang:2015aia,Sahay:2017hlq,Witten:2020ert,Wei:2020kra,Su:2019gby,Wang:2020pmb,Yan:2021uzw,Yerra:2021hnh,Zhao:2020nrx}. In general, these studies are based on the assumption that black holes can be treated as the thermodynamic entities. It can be shown that two stable thermodynamic phases emerge at certain temperature interval: the thermal AdS space phase and the large Schwarzschild-AdS black hole phase. By comparing the on-shell free energies of the two phases that can be computed from the gravitational action, it is found that there is a first order phase transition between the thermal AdS space and the AdS black hole at a certain critical temperature \cite{Hawking:1982dh}. The shortcoming of these studies based on the on-shell free energy is that only the thermodynamic stability can be analyzed while the dynamical transition process, i.e. the kinetics of the Hawking-Page phase transition, cannot be revealed.

A recent effort on this aspect is to employ the free energy landscape formalism \cite{FSW,FW,NG,JW} to study the dynamical processes of the Hawking-Page phase transition \cite{Li:2020khm} as well as the small/large RNAdS black hole phase transition \cite{Li:2020nsy,Li:2021vdp}. An essential concept is the order parameter of the macroscopic black hole state that represents the microscopic degrees of freedom of the black hole state. It is proposed that the black hole radius can be used to describe the on-shell black hole states as well as the off-shell black hole states. Here, the on-shell states refer to the macroscopic spacetime states that are solutions to the Einstein equations, while the off-shell state are the spacetime states generated by the thermal fluctuations during the phase transition process \cite{Li:2020khm}. For every off-shell spacetime state, the generalized free energy can be properly defined, and for the on-shell state, the generalized free energy can be reduced to the on-shell free energy which coincides with the results computed from the Einstein-Hilbert action. All the spacetime states compose a canonical ensemble, and the probability of the system in a specific state is then determined by the Boltzmann law $p\sim e^{-\beta G}$, where $\beta$ is the inverse temperature of the canonical ensemble and $G$ is the generalized free energy. Under the thermal fluctuations, the order parameter changes continuously, and the generalized free energy as the function of the order parameter has the shape of single well or double well depending on the ensemble temperature \cite{Li:2020khm,Li:2020nsy,Li:2021vdp}. This is the free energy landscape formalism of black hole phase transition.

In consequence, two benefits can be directly obtained from the landscape description. The first is that one can easily read the thermodynamic stable state from the landscape topography and analyze the black hole phase transition. The other is that the dynamical process and the kinetics of the black hole phase transition can be studied by assuming that the transition process is stochastic crossing the potential barrier representing the intermediate state of the phase transition. In this aspect, we assume that the dynamics is governed by three types of interactions \cite{Li:2021vdp}. The first comes from the thermodynamic driving force emergent from the interactions among all the microscopic degrees of freedom of the black hole. The free energy landscape plays the role of an effective potential. The second force is the effective friction along the order parameter. It can be interpreted as the interaction or the dissipation of the microscopic degrees of freedom from the effective heat bath acting on the black hole collective order parameter. The third force represents the stochastic force that comes from the microscopic degrees of freedom of the effective heat bath. Under these assumptions, the Langevin equation or the equivalent Fokker-Planck equation are employed to calculate the kinetic time and reveal the dynamical properties of the black hole phase transitions \cite{Wei:2020rcd,Li:2020spm,Wei:2021bwy,Cai:2021sag,Lan:2021crt,Li:2021zep,Yang:2021nwd,Mo:2021jff,Kumara:2021hlt,Li:2021tpu,Liu:2021lmr,Xu:2021usl,Du:2021cxs}. In \cite{Li:2021vdp}, the possibility of probing the black hole microstructure from the kinetic turnover of the small/large RNAdS black hole phase transition was discussed.

It should be pointed out that in the previous studies, the dynamics of the black hole phase transition is assumed to be Markovian, i.e. the correlation time of the effective thermal bath is assumed to be much shorter than the oscillating time of the spacetime state in the potential well on the free energy landscape. If this condition is not valid, then the non-Markovian model of the black hole phase transition is required. As pointed out in \cite{Li:2021vdp}, if the effective thermal bath is Markovian, the friction kernel originated from the interactions of the microscopic degrees of freedom of the effective heat bath acting on the order parameter is instantaneous. Since the non-Markovian bath has the memory, which couples the present dynamics to the past states, the non-Markovian effects can be described by the time dependent friction kernel \cite{Kubo:1985,Nitzan:2006,Zwanzig:1961}. It would be of great interest in an investigation of the non-Markovian effects on the kinetics of black hole phase transitions.

In the present work, we will consider the non-Markovian effects on the kinetics of the Hawking-Page phase transition in terms of the generalized Langevin equation where the non-Markovian effects are reflected by the time dependent friction kernels \cite{Kubo:1985,Nitzan:2006,Zwanzig:1961}. More specifically, we will employ the Grote-Hynes model in chemical physics for memory effect \cite{Grote:1980II} to discuss the dynamical transition process from the SAdS black hole to the thermal AdS space in Hawking-Page phase transition. We use the transition rate to quantify the dynamics of the Hawking-Page phase transition. We give the analytical expression of the transition rate and discuss the corresponding numerical results. As the concrete examples, we study the effects of the exponentially decay friction kernel and the oscillatory friction kernel on the kinetics of Hawking-Page phase transition. For the exponentially decayed friction, the non-Markovian effects promote the transition process, and for the oscillatory friction, increasing the oscillating frequency also speeds up the transition process.   

This paper is arranged as follows. In section \ref{Free_energy}, we briefly review the free energy landscape formalism of the Hawking-Page phase transition. In section \ref{Non_Markovian_model}, we discuss the basic setup of the non-Markovian model for black hole phase transition and give a brief derivation of the transition rate for the Hawking-Page phase transition. In section \ref{Num_results}, the numerical results for the exponentially decay friction kernel and the oscillatory friction kernel are discussed in detail and the non-Markovian effects on the Hawking-Page phase transition are revealed by comparing the Markovian model. This conclusions and discussions are presented in section \ref{Con_dis}.

\section{Free energy landscape of Hawking-Page phase transition}
\label{Free_energy}

In this section, we briefly review the free energy landscape description of Hawking-Page phase transition. Without loss of generality, we consider the four dimensional Schwarzschild-AdS (SAdS) black hole, the metric of which is given by \cite{Hawking:1982dh} 
\begin{eqnarray}\label{SAdS}
ds^2=-\left(1-\frac{2M}{r}+\frac{r^2}{L^2}\right)dt^2+
\left(1-\frac{2M}{r}+\frac{r^2}{L^2}\right)^{-1}dr^2+r^2d\Omega_2^2\;,
\end{eqnarray}
where $M$ and $L$ are the black hole mass and the AdS curvature radius. When $M=0$, the metric given by (\ref{SAdS}) gives the metric of pure AdS spacetime in four dimensions. The mass, the Hawking temperature, and the Bekenstein-Hawking entropy of SAdS black hole are respectively given by
\begin{eqnarray}
M&=&\frac{r_+}{2}\left(1+\frac{r_+^2}{L^2}\right) \;,
\\
T_H&=&\frac{1}{4\pi r_+}\left(1+\frac{3r_+^2}{L^2}\right)\;,\\
S&=&\pi r_+^2\;,
\end{eqnarray}
where $r_+$ is the black hole radius. The black hole radius is determined by the equation $1-\frac{2M}{r}+\frac{r^2}{L^2}=0$ and given by the largest root of this equation.

As discussed in \cite{Li:2020khm}, we consider the canonical ensemble at the specific temperature $T$ composed of a series of black hole spacetimes with an arbitrary horizon radius. Besides the thermal AdS space and the small/large SAdS black holes, which are the solutions to the Einstein equations, the assumed black holes with arbitrary horizon radii are the fluctuating black holes created by the thermal fluctuations. We call the black hole spacetime in the ensemble as the black hole state. According to the ensemble theory, every black hole state has the specific probability determined by the generalized free energy function.

It is well known that the on-shell Gibbs free energy for the on-shell solutions to the Einstein equation can be given by the thermodynamic relationship $G=M-T_HS$ or calculated directly from the Euclidean action \cite{Gibbons:1976ue}. In order to quantify the free energy landscape, we need to specify every spacetime state in the ensemble a Gibbs free energy. We generalize the on-shell free energy to the off-shell Gibbs free energy by replacing the Hawking temperature $T_H$ with the ensemble temperature $T$, which is explicitly given as follows \cite{York:1986it,Spallucci:2013osa,Andre:2020czm,Andre:2021ctu}
\begin{eqnarray}
G=M-TS=\frac{r_+}{2}\left(1+\frac{r_+^2}{L^2}\right)-\pi T r_+^2\;.
\end{eqnarray}
Note that the black hole radius is considered as the order parameter emerged from the underlying microscopic degrees of freedom of the system. The off-shell free energy is expressed as a continuous function of the order parameter and ensemble temperature.

\begin{figure}
  \centering
  \includegraphics[width=8cm]{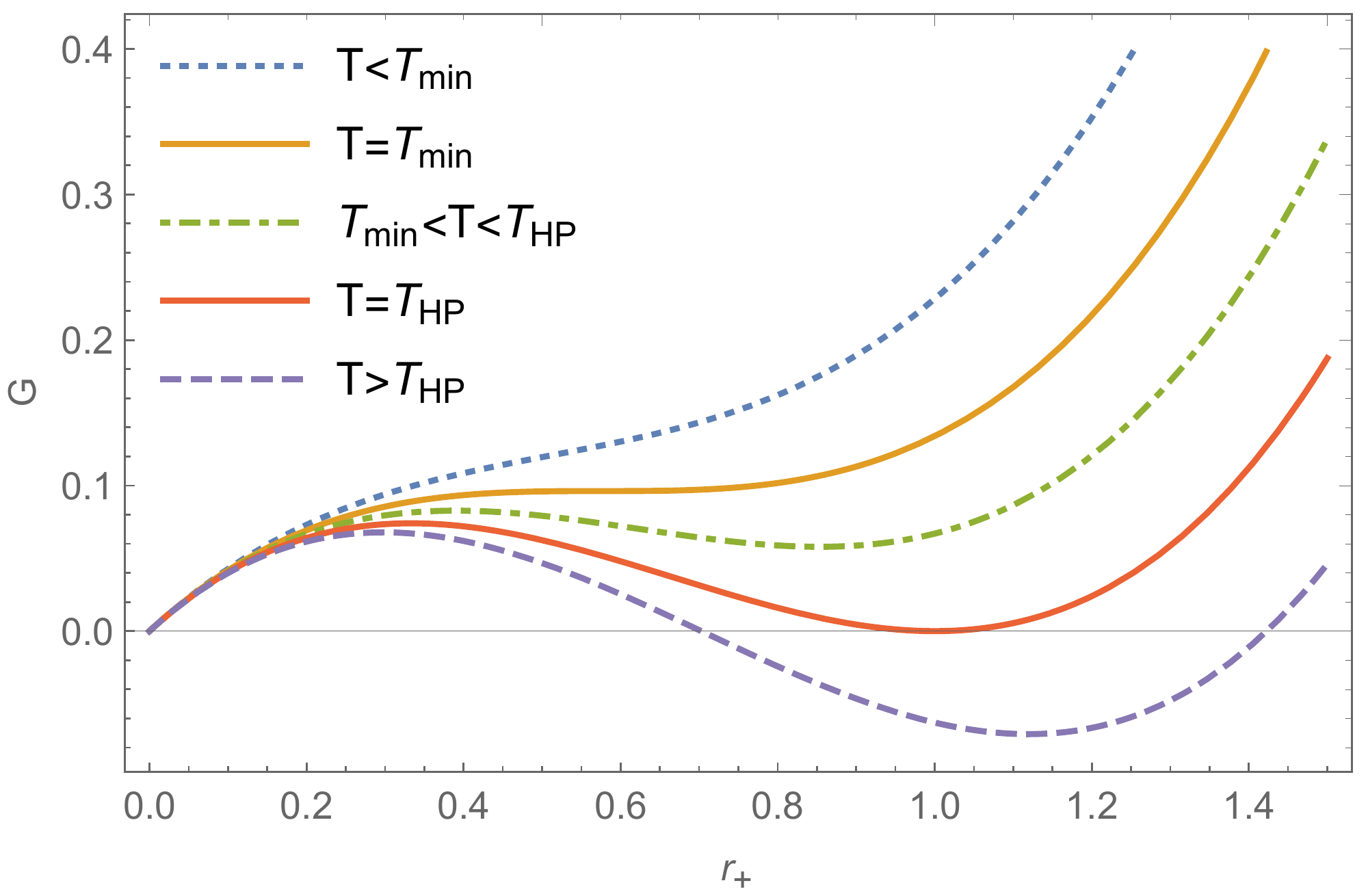}
  \caption{Free energy landscapes for Hawking-Page phase transition. The generalized Gibbs free energy is plotted as the function of black hole radius $r_+$ at different temperature. In discussing Hawking-Page phase transition, we set the AdS curvature radius $L$ as unity. }
  \label{GibbsPlot}
\end{figure}

In Figure \ref{GibbsPlot}, the free energy landscapes for the Hawking-Page phase transition are plotted for different ensemble temperatures. It can be seen that the free energy landscape for the Hawking-Page phase transition is only modulated by the ensemble temperature. In these plots, the origin $r_+=0$ represents the pure radiation phase or thermal AdS space. When $T<T_{min}=\frac{\sqrt{3}}{2\pi L}$, there is only one global minimum on the free energy landscape at the origin. At $T=T_{min}$, Gibbs free energy landscape exhibits an inflection point at $r_+=L/\sqrt{3}$. From this temperature on, the two black hole phases emerge (large and small black holes) with radii given by
\begin{eqnarray}\label{rls}
r_{l,s}=\frac{T}{2\pi T_{min}^2}\left(1\pm\sqrt{1-\frac{T_{min}^2}{T^2}}\right)\;.
\end{eqnarray}
The small black hole phase is unstable while the large black hole phase can be stable or unstable depending on the temperature. 

The stability can be analyzed by using the principle that the global or the local minimum on the free energy landscape corresponds to the global or the local stable phase. It can be easily read off from the free energy landscapes that below the Hawking-Page critical temperature $T=T_{HP}=\frac{1}{\pi L}$, the thermal AdS phase is thermodynamically stable, and above the critical temperature, the large black hole phase is thermodynamically stable. At the critical temperature, both the thermal AdS space phase and the large black hole phase are stable with equal free energy basin depth. Since there is a discontinuous change in the order parameter from the thermal AdS phase $r_+=0$ to the large black hole phase $r_+=r_l$ separated by a free energy barrier at the Hawking-Page critical temperature, the associated derivative of the free energy function will be discontinuous, which is a signature of the first order phase transition. It should be noted that besides the thermal AdS space and the small and large SAdS black holes, there are fluctuating black holes on the free energy landscape. The fluctuating black holes represent the intermediate states of the phase transition process.

\section{Non-Markovian dynamics of the black hole phase transition} 
\label{Non_Markovian_model}

In this section, we introduce the basic concepts and models of phase transition involving the non-Markovian effects. The application of the non-Markovian model to the black hole phase transition is particularly addressed.  

\subsection{Generalized Langevin equation}

In contrast to the previous Markovian model of the black hole phase transition on the free energy landscape, the non-Markovian dynamics can be described by the generalized Langevin equation complemented by the free energy potential landscape $G(r)$ \cite{Mori:1965,Kubo:1966}
\begin{eqnarray}\label{Non_Mar_Eq}
&&\frac{dr}{dt}=v\;,\nonumber\\
&&\frac{dv}{dt}=-\frac{dG(r)}{dr}-\int_{0}^{t} \zeta(\tau) v(t-\tau) d\tau
+\eta(t)\;,
\end{eqnarray}
where the color noise $\eta(t)$ with zero mean and the dissipation kernel $\zeta(t)$ are related by the second fluctuation-dissipation theorem 
\begin{eqnarray}
\zeta(|t|)=\frac{1}{k_B T} \langle \eta(0)\eta(t)\rangle\;.
\end{eqnarray}
In Eq.(\ref{Non_Mar_Eq}), we have abbreviated the order parameter $r_+$ to $r$ for simplicity. As noted, the Langevin equation with the Markovian dynamics in \cite{Li:2021vdp} can be derived from Eq.(\ref{Non_Mar_Eq}) by taking $\zeta(t)=\zeta\delta(t)$. The non-Markovian effects is reflected by the time dependent friction kernel $\zeta(t)$. The generalized Langevin equation can be viewed as the effective equation that describes the evolution of the black hole order parameter at the macroscopic emergent level. It is an integro-differential equation that is hard to deal with directly.

In writing down the generalized Langevin equation (\ref{Non_Mar_Eq}), we have assumed that the effective stochastic dynamics describing the evolution along the order parameter is determined by the three forces specified below, which is the analogy to the assumptions made in the Markovian model \cite{Li:2021vdp}. The first term is the thermodynamic driving force emergent from the interactions among all the microscopic degrees of freedom of the black hole. The free energy landscape plays the role of an effective potential. The second force is the effective friction along the order parameter. It is interpreted as the interaction or the dissipation of the microscopic degrees of freedom from the effective heat bath acting on the black hole collective order parameter. In formulating the non-Markovian model, the correlation time of the thermal bath is assumed to be comparable or even longer than the oscillating time of the black hole state at the potential well on the free energy landscape. Here, the oscillating time of the black hole state means the fluctuation or relaxation time scale of the black hole state at the potential well on the free energy landscape. The non-Markovian effect can be represented by the time dependent friction \cite{Kubo:1985,Nitzan:2006,Zwanzig:1961}. The third force represents the stochastic force that comes from the microscopic degrees of freedom of the effective heat bath on the macroscopic order parameter. In the non-Markovian model, the noise is colored and satisfies the second fluctuation-dissipation theorem.

\subsection{Transition rate with non-Markovian effects}

\begin{figure}
  \centering
  \includegraphics[width=8cm]{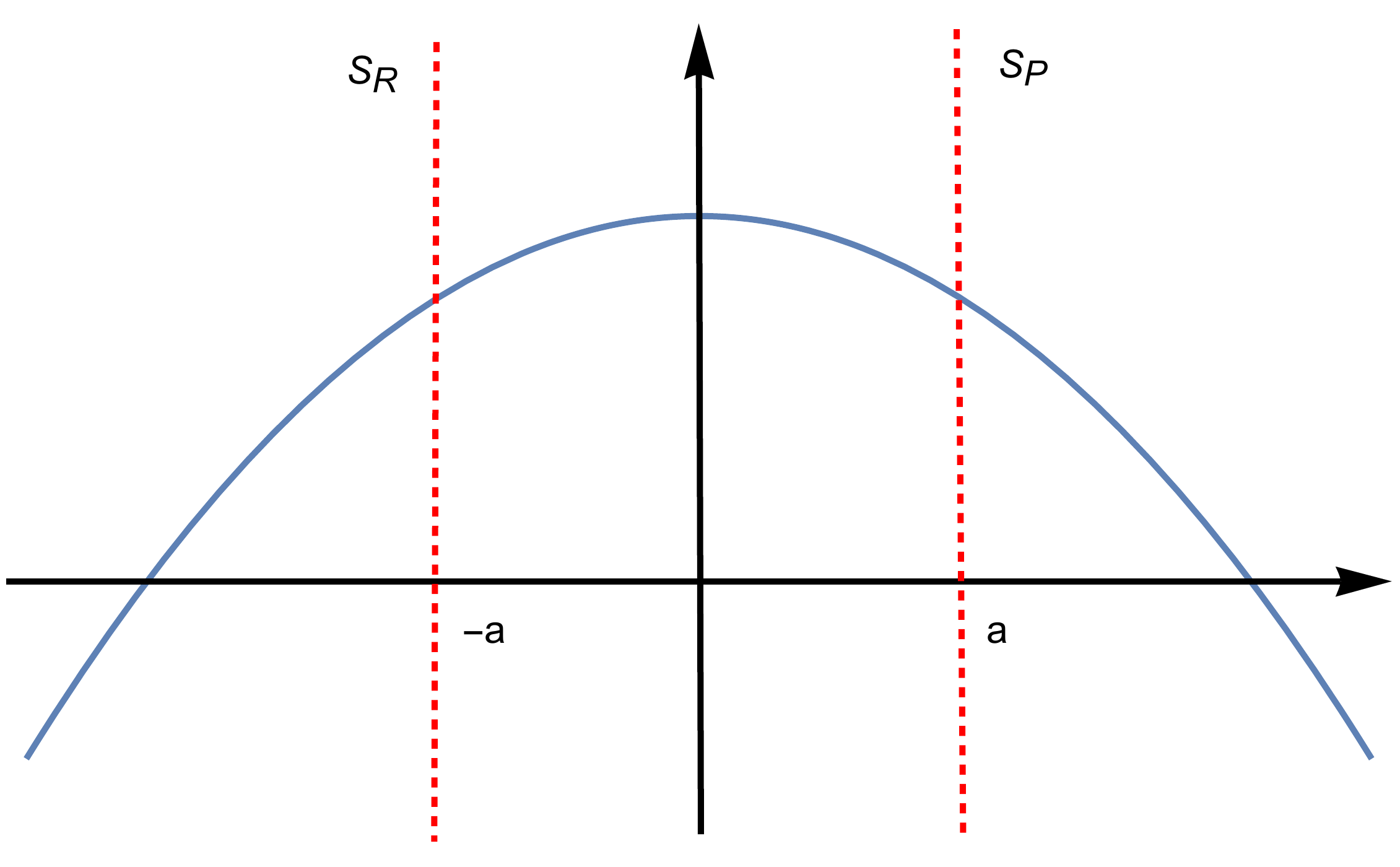}
  \caption{Schematic plot of barrier crossing model of black hole phase transition with non-Markovian effects. }
  \label{Scetch}
\end{figure}

In the Markovian model of the black hole phase transition, the transition process from one black hole state to another black hole state is treated as stochastic crossing the potential barrier on the free energy landscape under the thermal fluctuations. The dynamics of the effective bath do not play a very significant role in the barrier crossing. As mentioned in the introduction, the Markovian assumption breaks down if the bath relaxes on a similar or slower timescale as the timescale of the black hole in the well on the free energy landscape. In such a case, the non-Markovian effect should be taken into account. The corresponding transition rate that reflects the kinetics of the transition process was first studied in the 1980s in the chemical physics context \cite{Grote:1980II}. Grote and Hynes used the generalized Langevin equation to derive the transition rate taking the non-Markovian effects into account \cite{Grote:1980II}. The schematic diagram is plotted in Figure \ref{Scetch}. Only the dynamics near the top of the barrier is concerned. Without loss of generality, we set the top of the barrier as $r=0$ temporally in this subsection. $S_R$ and $S_P$ are two surfaces placed on the left side ($r=-a$) and the right side ($r=a$) of the barrier. These surface are placed far enough from the barrier region that the black hole state crossing the barrier to the left region or the right region through the surface $S_R$ or $S_P$ do not return the barrier region again. Grote and Hynes derived the transition rate of the barrier crossing from the surface $S_R$ to the surface $S_P$ \cite{Grote:1980II}. In fact, they consider the switching process in the double well case, i.e., there is a potential well on each side of the barrier. We will derive the transition rate from the left well to the right well firstly and apply the result to discuss the kinetics of the Hawking-Page phase transition in the next subsection.

First, let us define the transition rate of this switching process. The transition rate is defined as the correlation function of the right-going flux passing through $S_R$ at time zero, $j(S_R,0)$, with the right-going flux passing through $S_P$, $j(S_P,t)$, at all later times. This quantity describes the kinetics of transition process. Accordingly, the transition rate can be explicitly given by \cite{Grote:1980II,Kohen:1995}
\begin{eqnarray}
\kappa=\int_0^\infty dt \langle j(S_R,0)j(S_P,t) \rangle\;,
\end{eqnarray}
where the average is taken over an equilibrium distribution. 

Define $\rho(r,v,t|r_0,v_0)$ as the probability that the black hole state in the canonical ensemble will have the order parameter $r$ and its velocity $v$ at time $t$, given that it had the order parameter $r_0$ and $v_0$ at time $t=0$. In terms of the probability distribution, one can rewrite the transition rate as 
\begin{eqnarray}
\kappa=\int_0^\infty dt \int dr dv \int dr_0 dv_0 
\phi_{eq}(v_0)\psi_{eq}(r_0) v_0 \delta(r_0+a)  
v \delta(r-a) \rho(r,v,t|r_0,v_0)\;.
\end{eqnarray}
Here, the $\delta-$functions take the initial conditions into account. $\phi_{eq}(v_0)$ is the Maxwell equilibrium distribution and $\psi_{eq}(r_0)$ is the equilibrium distribution in the left well on the landscape. They are given by 
\begin{eqnarray}
\phi_{eq}(v_0)&=&\frac{1}{\sqrt{2\pi k_B T}}\exp[-v_0^2/2k_B T]\;,
\nonumber\\
\psi_{eq}(r_0)&=&\frac{\omega_L}{\sqrt{2\pi k_B T}}\exp[-\beta G(r_0)]\;,\nonumber
\end{eqnarray}
where $\omega_L=\sqrt{G''(r)}|_{r=r_L}$ is the oscillating frequency of the black hole state in the left potential well ($r=r_L$) on the free energy landscape.

The transition rate can be simplified as 
\begin{eqnarray}\label{kappa_sim}
\kappa=\psi_{eq}(-a)\int_0^\infty dt \int  dv \int  dv_0 \phi_{eq}(v_0) v_0  
v \rho(a,v,t|r_0,v_0)\;, 
\end{eqnarray}
where 
\begin{eqnarray}
\psi_{eq}(-a)&=&\sqrt{\frac{\beta\omega_L^2}{2\pi}}e^{-\beta G(-a)}\nonumber\\
&\cong&\sqrt{\frac{\beta\omega_L^2}{2\pi}} e^{-\beta W +\frac{\beta}{2}\omega_M^2 a^2}\;. 
\end{eqnarray}
In deriving the expression for $\psi_{eq}(r_0=-a)$, we have also used the quadratic approximation of the free energy function. Here, $W=G(0)-G(r_L)$ is the barrier height between the black hole state on the top of the landscape ($r=0$) and the black hole state in the left well of the landscape ($r=r_L$), and $\omega_M=\sqrt{|G''(r)|}|_{r=0}$ is the oscillating frequency of the black hole state on the top of the landscape.

In Eq.(\ref{kappa_sim}), the value of $r_0$ in the probability distribution function $\rho(a,v,t|r_0,v_0)$ should be taken as $r_0=-a$. One should recognize that 
\begin{eqnarray}
\int_0^\infty dt \int  dv v \rho(a,v,t|r_0,v_0)
\end{eqnarray}
is the integrated net flux across $S_P$, which can be rewritten as 
\begin{eqnarray}\label{net_flux}
\lim_{t\rightarrow \infty} \int_{a}^{\infty} dr \int dv \rho(r,v,t|r_0,v_0)\;.  
\end{eqnarray}

To derive the transition rate, we should firstly obtain the probability distribution function $\rho(r,v,t|r_0,v_0)$. Since the random force $\eta(t)$ is assumed to be Gaussian, the probability distribution is shown to be also Gaussian, which is given by \cite{Kohen:1995,Adelman:1976} 
\begin{eqnarray}
\rho(r,v,t|r_0,v_0)=\frac{1}{\pi}\left|\det{Q^{-1}} \right|^{1/2}
\exp(-y^T\cdot Q^{-1}\cdot y)\;, 
\end{eqnarray}
where 
\begin{eqnarray}
y^T&=&(r(t)-\langle r(t) \rangle,v(t)-\langle v(t) \rangle)\;,
\nonumber\\
Q_{ij}&=&2\langle y_i(t) y_j(t) \rangle\;. 
\end{eqnarray}
The averages in the above equations should be taken over an evolving swarm of trajectories that accounts the initial conditions $r(0)=r_0=-a$ and $v(0)=v_0$.

Inserting the probability distribution into Eq.(\ref{net_flux}) and performing the integrals, one can get
\begin{eqnarray}
 \int_{a}^{\infty} dr \int dv \rho(r,v,t|r_0,v_0)&=&\int_{a}^{\infty}
 dr (\pi Q_{11})^{-1/2} \exp(-y_1^2/Q_{11})\nonumber\\
 &=&\frac{1}{2}\textrm{erfc} \left[ \frac{a-\langle r\rangle}{\sqrt{Q_{11}}} \right]\;,
\end{eqnarray}
where $\textrm{erfc}$ is the complementary error function.

To proceed, we need the average of $r(t)$, which can be easily obtained by solving the generalized Langevin equation in terms of the Laplace transformation. The generalized Langevin equation near the top of the barrier can be approximated by 
\begin{eqnarray}
\frac{d^2 r}{dt^2}=\omega_M^2 r-\int_{0}^{t} \zeta(\tau) v(t-\tau) d\tau
+\eta(t)\;,
\end{eqnarray}
where we have used the quadratic approximation of the free energy function. The Laplace transformation of the generalized Langevin equation gives
\begin{eqnarray}
\lambda^2 \hat{r}(\lambda) +\lambda a -v_0=\omega_M^2 \hat{r}(\lambda) -\hat{\zeta}(\lambda)(\lambda \hat{r}(\lambda)- a) 
+\hat{\eta}(\lambda)\;,
\end{eqnarray}
where the initial conditions $r(0)=-a$ and $v(0)=v_0$ are used. Note that the quantities with hat are the Laplace transforms of the corresponding quantities. By solving this equation, one can get 
\begin{eqnarray}\label{rt_eq}
r(t)=-\chi_r(t)a+\chi_v v_0+\int_0^t \chi_v(\tau) \eta(t-\tau)d\tau\;,
\end{eqnarray}
where 
\begin{eqnarray}
\chi_v(t)&=&\mathcal{L}^{-1}[\lambda^2+\lambda\hat{\zeta}(\lambda)-\omega_M^2]^{-1}\;,\\
\chi_r(t)&=&1+\omega_M^2 \int_0^t \chi_v(\tau) d\tau\;.
\end{eqnarray}
Taking the average of Eq.(\ref{rt_eq}), one can get 
\begin{eqnarray}
\langle r\rangle=-\chi_r(t)a+\chi_v v_0\;.
\end{eqnarray}
In deriving this result, we used the fact that the average of the random force is zero.

With these results, performing the integral over $v_0$, one can get the transition rate as 
\begin{eqnarray}
\kappa= \frac{\omega_L}{\pi\sqrt{2\beta}}e^{-\beta W} 
\lim_{t\rightarrow \infty} \left\{ \chi_v \phi 
\exp\left[ \frac{1}{2}\beta a^2 \omega_m^2-a^2\phi^2(1+\chi_r)^2   \right] \right\}\;,
\end{eqnarray}
where $\phi^{-1}=\sqrt{Q_{11}+2\chi_v^2/\beta}$.

Now, we have to analyze the late time behaviors of $\chi_r(t)$ and $\chi_v(t)$. Note that the Laplace transform of $\chi_v(t)$ is 
\begin{eqnarray}
\hat{\chi}_v(\lambda)=\frac{1}{\lambda^2+\lambda\hat{\zeta}(\lambda)-\omega_M^2}\;.
\end{eqnarray}
The equation $\lambda^2+\lambda\hat{\zeta}(\lambda)-\omega_M^2=0$ has one root $\lambda_r$ in the interval $0<\lambda<\omega_M$. Therefore, $\chi_v(\lambda)$ is divergent at $\lambda=\lambda_r$ and the late time behavior of $\chi_v(t)$ is determined by 
\begin{eqnarray}
\chi_v(t)\sim e^{\lambda_r t}\;. 
\end{eqnarray}
In consequence, the late time behavior of $\chi_r(t)$ is given by 
\begin{eqnarray}
\chi_r(t)\sim \frac{\omega_M^2}{\lambda_r} e^{\lambda_r t}\;. 
\end{eqnarray}
Therefore, the late time behaviors of $\chi_r(t)$ and $\chi_v(t)$ must be exponentially diverging in time.

Since the transition rate is irrelevant to the selection of the surface $S_R$ or $S_p$, $\kappa$ is independent of $a$. Therefore, by using the divergent behaviors of $\chi_r(t)$ and $\chi_v(t)$, we can conclude that  
\begin{eqnarray}
\lim_{t\rightarrow \infty} \phi=\lim_{t\rightarrow \infty} \left(\frac{\beta \omega_M^2}{2}\right)^{1/2} \frac{1}{\chi_r}\;.
\end{eqnarray}
The transition rate simplifies to 
\begin{eqnarray}
\kappa&=&\frac{\omega_L\omega_M}{2\pi}e^{-\beta W} \lim_{t\rightarrow \infty} \frac{\chi_v(t)}{\chi_r(t)}\nonumber\\
&=&\frac{\omega_L}{2\pi}e^{-\beta W} \lim_{t\rightarrow \infty} \frac{\chi_v(t)}{\omega_M\int_0^t\chi_v(\tau)d\tau}\;.
\end{eqnarray}

Substituting the late time behavior of $\chi_v(t)$ into the above equation, we can finally get the transition rate as 
\begin{eqnarray}\label{tran_rate}
\kappa=\frac{\lambda_r}{\omega_M}\frac{\omega_L}{2\pi}e^{-\beta W}\;,
\end{eqnarray}
where $\lambda_r$ is the largest positive root of the equation 
\begin{eqnarray}\label{freq_eq}
\lambda=\frac{\omega_M^2}{\lambda+\hat{\zeta}(\lambda)}\;.
\end{eqnarray}

This is the analytical result of the transition rate from the black hole state in the left well to the black hole state in the right well. It is related to the barrier height $W=G(0)-G(r_L)$, the oscillating frequencies $\omega_L=\sqrt{G''(r)}|_{r=r_L}$ and $\omega_M=\sqrt{|G''(r)|}|_{r=0}$, as well as the positive root of Eq.(\ref{freq_eq}). One can refer to \cite{Peters:2017} for a comprehensive review on the reaction rate theory.

\subsection{The rate of Hawking-Page phase transition}

We consider the transition process from the SAdS black hole state (represented by the right potential well on the free energy landscape in Figure \ref{GibbsPlot}) to the thermal AdS space state (represented by the origin of the landscape in Figure \ref{GibbsPlot}). Compared with the schematic plot in Figure \ref{Scetch}, the small SAdS black hole state is represented by the point $r=0$ at the top of the potential barrier in Figure \ref{Scetch} and the large black hole state is represented by the right well on the landscape. Therefore, the analytical results in the last subsection can be transformed into the transition rate of the process from the SAdS black hole state to the thermal AdS space state in terms of the following replacement 
\begin{eqnarray}
\omega_L&\rightarrow& \omega_l=\sqrt{G''(r)}|_{r=r_l}\;,\nonumber\\
\omega_M&\rightarrow&
\omega_s=\sqrt{|G''(r)|}|_{r=r_s}\;,\\
W&\rightarrow& W=G(r_s)-G(r_l)\;,\nonumber
\end{eqnarray}
where $r_{s/l}$ is the radius of the small/large SAdS black hole given in Eq.(\ref{rls}). The transition rate of the process from the SAdS black hole state to the thermal AdS space state is then given by 
\begin{eqnarray}\label{tran_rate_HP}
\kappa=\frac{\lambda_r}{\omega_s}\frac{\omega_l}{2\pi}e^{-\beta W}\;,
\end{eqnarray}
where $\lambda_r$ is the largest positive root of the equation 
\begin{eqnarray}\label{freq_eq_HP}
\lambda=\frac{\omega_s^2}{\lambda+\hat{\zeta}(\lambda)}\;, 
\end{eqnarray}
with $\hat{\zeta}(\lambda)$ being the Laplace transform of the friction kernel 
\begin{eqnarray}
\hat{\zeta}(\lambda)=\int_0^\infty \zeta(\tau) e^{-\lambda\tau} d\tau\;.
\end{eqnarray}

Now, let discuss the analytical result. By solving the quadratic equation (\ref{freq_eq_HP}) for $\lambda$ as the function of $\omega_s$, the transition rate for the non-Markovian friction can be written in the form of
\begin{eqnarray}\label{kappa_new}
\kappa=\left[\sqrt{1+\left(\frac{\hat{\zeta}(\lambda_r)}{2\omega_s}\right)^2}-\frac{\hat{\zeta}(\lambda_r)}{2\omega_s} \right]\frac{\omega_l}{2\pi}e^{-\beta W}\;.
\end{eqnarray}
This expression of the transition rate is exactly the transition rate derived by kramers for the intermediate friction case, except that the Markovian friction has been replaced by the Laplace transform of the non-Markovian friction at the frequency $\lambda_r$. In fact, if we consider the friction of the delta function
\begin{eqnarray}
\zeta(t)=\zeta \delta(t)\;,
\end{eqnarray}
the Laplace transformation gives 
\begin{eqnarray}
\hat{\zeta}(\lambda)=\zeta\;.
\end{eqnarray}
As mentioned, under this assumption, the non-Markovian Langevin equation reduces to the Markovian case. Therefore, we can get the transition rate expression for the Markovian friction as \cite{Li:2021vdp}
\begin{eqnarray}\label{kappa_Mar}
\kappa=\left[ \sqrt{1+\left(\frac{\zeta}{2\omega_s}\right)^2}-\frac{\zeta}{2\omega_s} \right]\frac{\omega_l}{2\pi}e^{-\beta W}\;.
\end{eqnarray}

Equations (\ref{tran_rate_HP}) and (\ref{freq_eq_HP}) show precisely how the non-Markovian friction influences the transition rate. The transition rate depends on the barrier height on the free energy landscape exponentially. The non-Markovian friction affects on the prefactor of the transition rate. In the next section, we will employ the analytical results to study the effects of the non-Markovian frictions on the kinetics of black hole phase transition.

\section{Effects of time dependent frictions on the kinetics of Hawking-Page phase transition}\label{Num_results}

In this section, we discuss the numerical results of the kinetics of Hawking-Page phase transition represented by the transition rate from the large SAdS black hole state to the thermal AdS space state. In the following, we will discuss two types of the friction kernels, i.e., the exponentially decayed friction, and the oscillatory friction. We will study the dependence of the transition rate on the ensemble temperature, the friction strength, decay coefficient, and oscillating frequency.

\subsection{Exponentially decayed friction kernel}

For the exponentially decayed friction, we consider the effects of the time scale of the friction kernel on the kinetics of Hawking-Page phase transition. The time scale of the friction kernel describes the correlation time of the effective thermal bath. In order to compare with Markovian dynamics, i.e. the delta function friction kernel, we consider the following exponentially decayed friction kernel
\begin{eqnarray}\label{exp_kernel}
\zeta(t) = \frac{\zeta}{\gamma} e^{-\frac{|t|}{\gamma}}\;,
\end{eqnarray}
where $\zeta$ represents the strength of the friction and $\gamma$ is the time scale of the friction kernel. When $\gamma\rightarrow 0$, the exponentially decayed friction is reduced to the delta function friction kernel discussed in the last subsection. 

For the exponentially decayed friction kernel, the Laplace transformation is given by \begin{eqnarray}\label{zeta_laplace}
\hat{\zeta}(\lambda)=\frac{\zeta}{1+\lambda\gamma}
\;.
\end{eqnarray}
The zero frequency value $\hat{\zeta}(0)=\zeta$ corresponds to the Markovian case while the non-Markovian case corresponds $\lambda_r\neq 0$. In order to obtain the Grote-Hynes frequency $\lambda_r$ in Eq.(\ref{tran_rate_HP}), one should combine equations (\ref{freq_eq_HP}) and (\ref{zeta_laplace}) to solve the largest positive root. Since $\lambda_r>0$ and $\gamma>0$, $\hat{\zeta}(\lambda_r)$ is generally smaller than $\zeta$. Compared the non-Markovian rate Eq.(\ref{tran_rate_HP}) with the Markovian rate Eq.(\ref{kappa_Mar}), one can conclude that for the exponentially decayed friction kernel, the non-Markovian effects generally promote the kinetics of the phase transition.

\begin{figure}
  \centering
  \includegraphics[width=8cm]{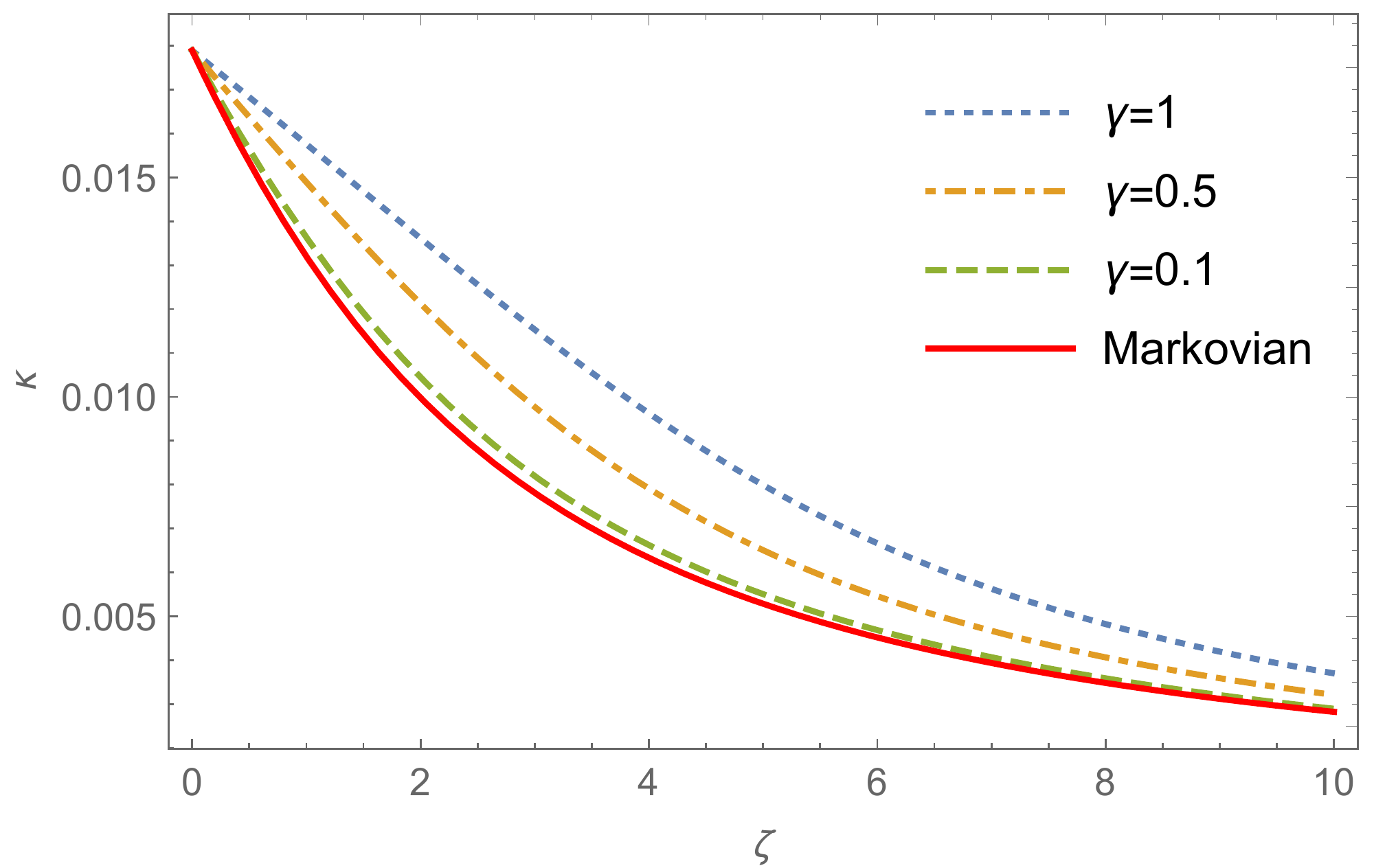}\\
  \caption{The dependence of the transition rate of the Hawking-Page phase transition from the large SAdS black hole state to the thermal AdS state on the friction coefficient $\zeta$ for the exponentially decayed friction kernel. In the plot, $L=1$, and $T=0.5$.    
  }\label{kappa_vs_zeta_exp_HP}
\end{figure}

In Figure \ref{kappa_vs_zeta_exp_HP}, we plot the dependence of the transition rate of the Hawking-Page phase transition from the large SAdS black hole state to the thermal AdS state on the friction coefficient $\zeta$ for the exponentially decayed friction kernel. It is shown that the non-Markovian effects promote the transition process. Increasing the correlation time $\gamma$ of the effective thermal bath will speed up the transition process, which is consistent with the above analyses. A physical intuitive understanding of the accelerated rate is that the non-Markovian effect introduces more time scales and therefore equivalently more degrees of freedom. So that the "short-cut" becomes possible. We have plotted the curves starting from $\zeta=0$. In fact, one should note that the Grote-Hynes theory is only valid in the intermediate-strong friction regime. The discussion of phase transition kinetics in the small friction regime is out of the scope of current work.

\begin{figure}
  \centering
  \includegraphics[width=8cm]{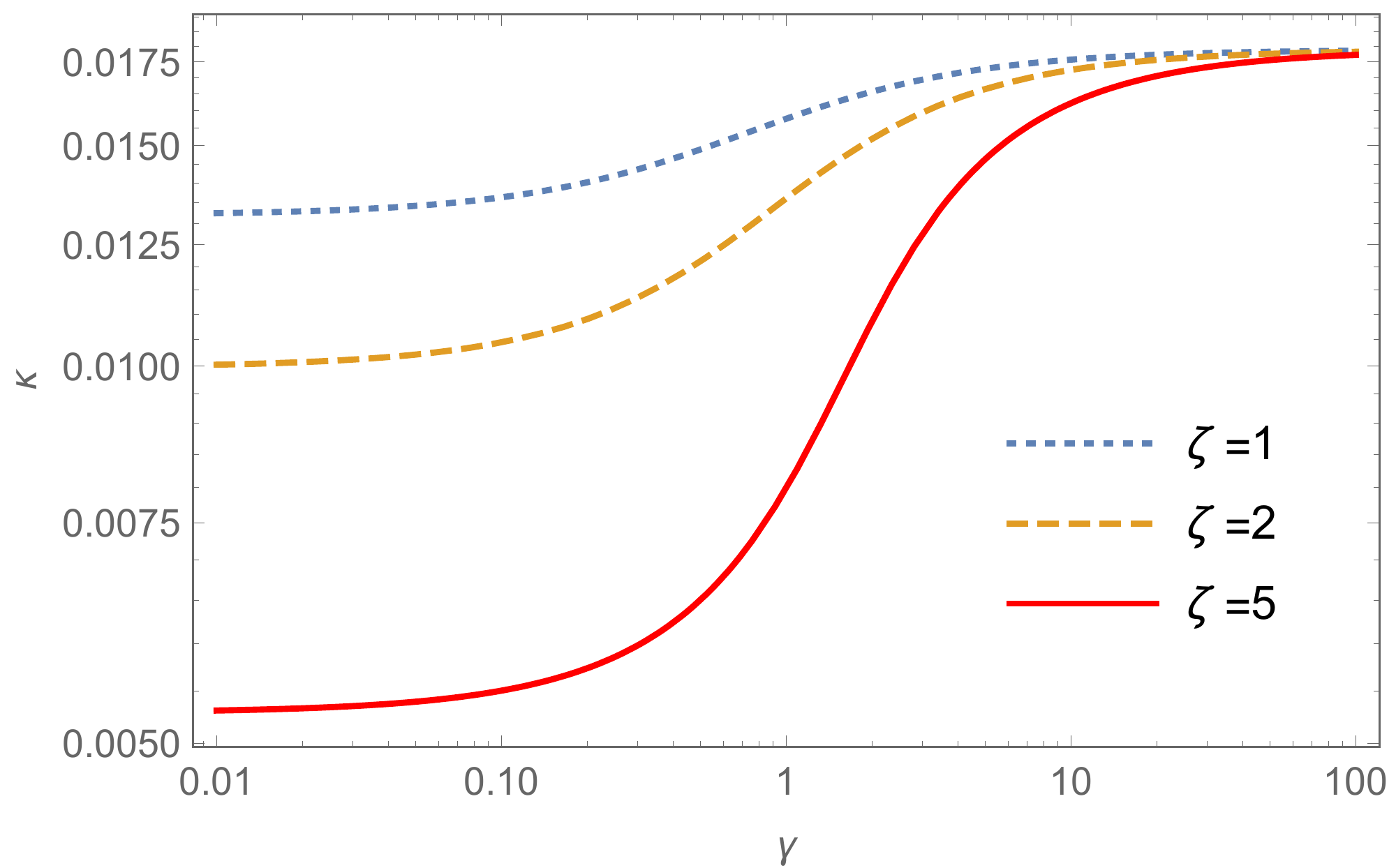}\\
  \caption{The dependence of the transition rate of the Hawking-Page phase transition from the large SAdS black hole state to the thermal AdS state on the decay coefficient $\gamma$ for the exponentially decayed friction kernel. In the plot, $L=1$, and $T=0.5$. }\label{kappa_vs_gamma_exp_HP}
\end{figure}

In Figure \ref{kappa_vs_gamma_exp_HP}, we plot the dependence of the transition rate of the Hawking-Page phase transition from the large SAdS black hole state to the thermal AdS state on the decay coefficient $\gamma$ for the exponentially decayed friction kernel. It is shown that the transition rate is the monotonic increasing function of the decay coefficient. In the Markovian limit $\gamma\rightarrow 0$, the transition rates for different friction coefficients have different limiting values, while in the limit $\gamma\rightarrow \infty$, their limiting values are same. For $\gamma\rightarrow 0$, the transition rate is determined by Eq.(\ref{kappa_Mar}) for the Markovian case. Different friction coefficient gives different transition rate. When $\gamma\rightarrow \infty$, $\hat{\zeta}(\lambda)\rightarrow 0$. The solution of Eq.(\ref{freq_eq_HP}) is given by $\lambda_r=\omega_m$, which is irrelevant to $\zeta$ and $\gamma$. In this case, the transition rate is an universal value which is given by $\kappa=\frac{\omega}{2\pi}e^{-\beta W}$.

\begin{figure}
  \centering
  \includegraphics[width=8cm]{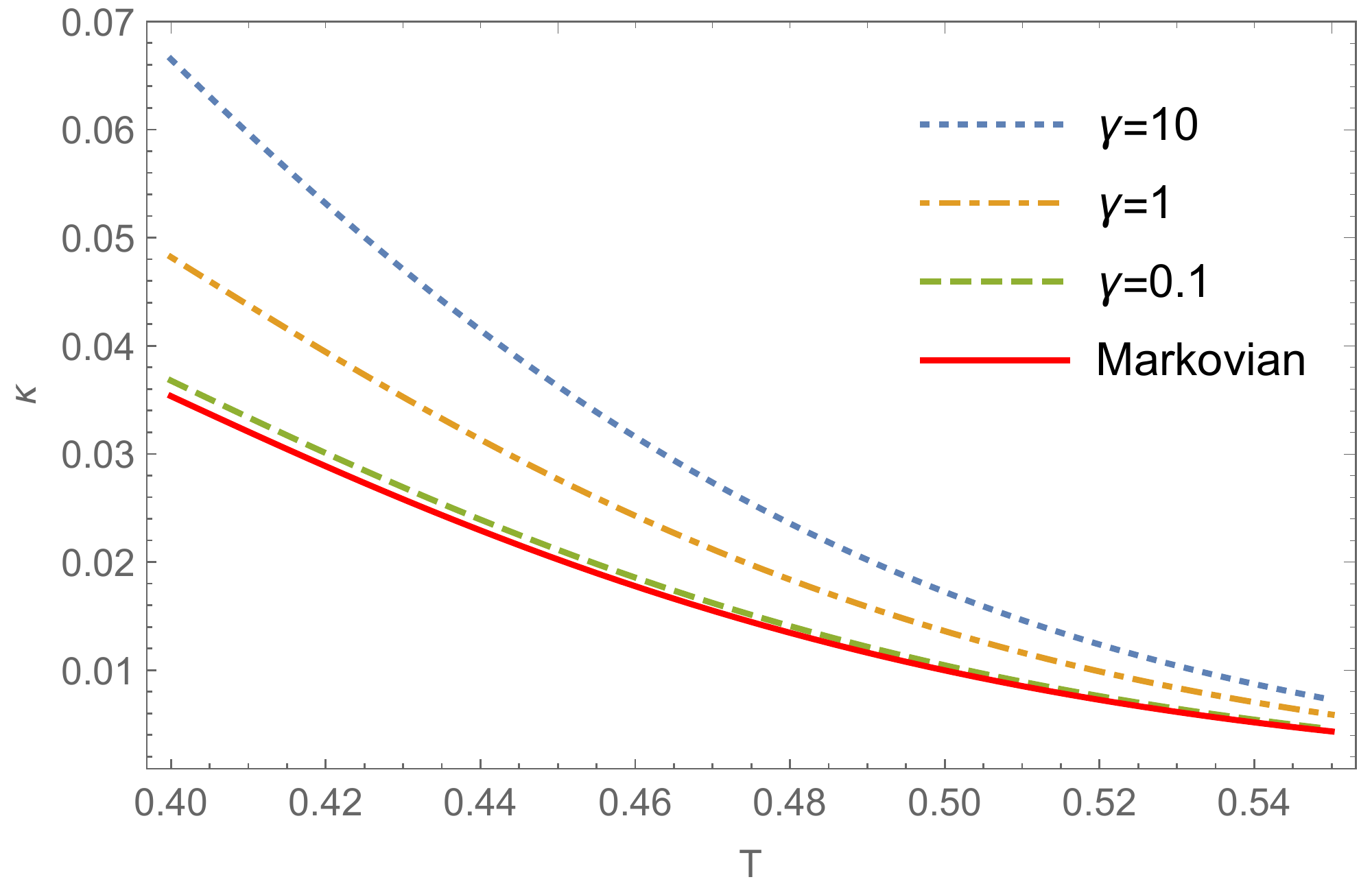}\\
  \caption{The dependence of the transition rate of the Hawking-Page phase transition from the large SAdS black hole state to the thermal AdS state on the ensemble temperature $T$ for the exponentially decayed friction kernel. In the plot, $L=1$ and $\zeta=2$. }\label{kappa_vs_T_exp_HP_zetafixed}
\end{figure}

In Figure \ref{kappa_vs_T_exp_HP_zetafixed}, we plot the dependence of the transition rate of the Hawking-Page phase transition from the large SAdS black hole state to the thermal AdS state on the ensemble temperature $T$ for the exponentially decayed friction kernel. It is shown that the transition rates for different decay parameters $\gamma$ are the monotonic decreasing functions of the ensemble temperature. The reason is that in the high barrier case, the kinetics is dominated by the barrier height. Increasing the temperature will lead to the increase of the barrier height between the small SAdS black hole state and the large SAdS black hole state. It can also be observed that large $\gamma$ results in large transition rate, i.e., the non-Markovian effects promote the transition process.

\begin{figure}
  \centering
  \includegraphics[width=8cm]{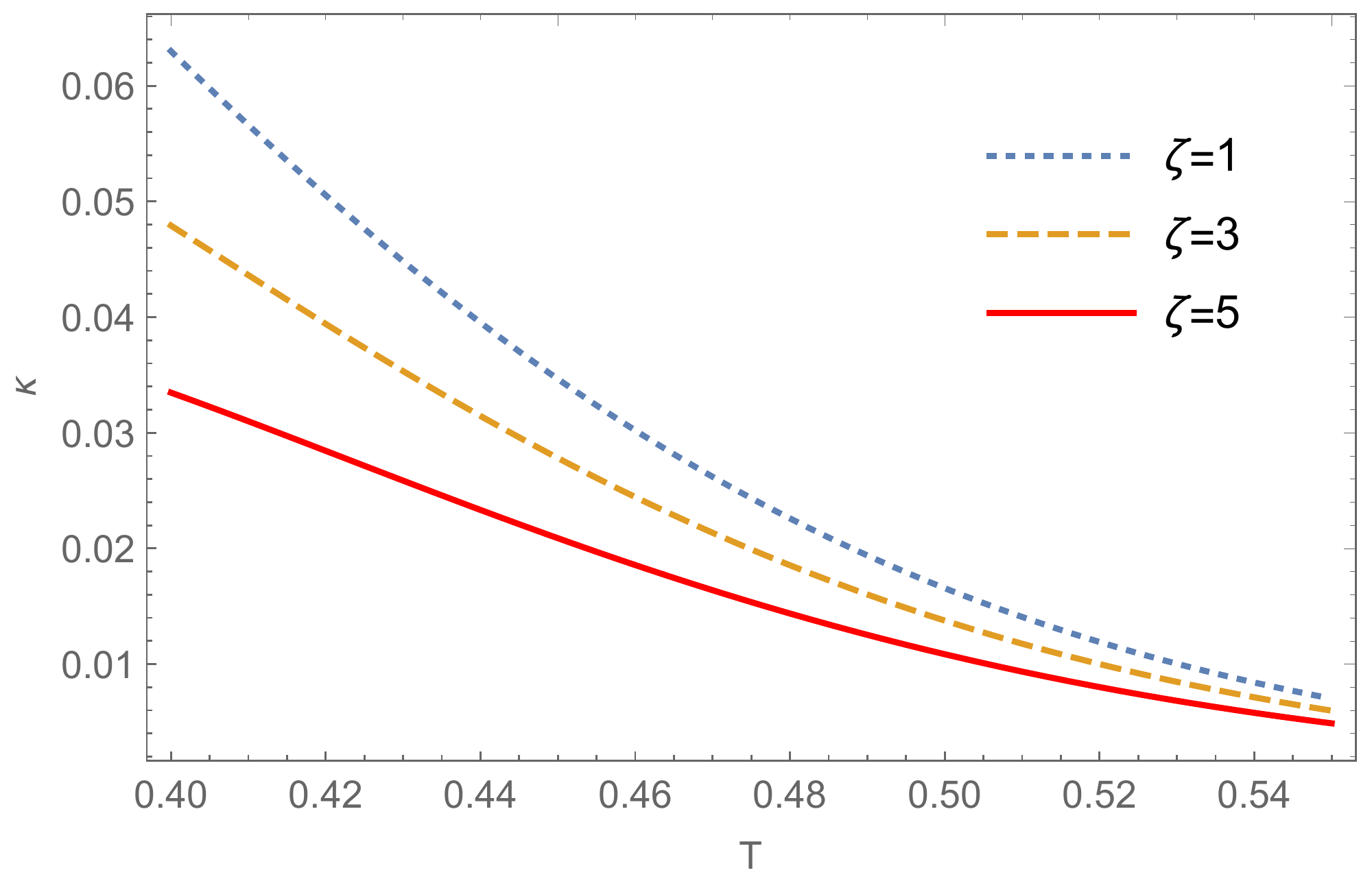}\\
  \caption{The dependence of the transition rate of the Hawking-Page phase transition from the large SAdS black hole state to the thermal AdS state on the ensemble temperature $T$ for the exponentially decayed friction kernel. In the plot, $L=1$ and $\gamma=2$. }\label{kappa_vs_T_exp_HP_gammafixed}
\end{figure}

In Figure \ref{kappa_vs_T_exp_HP_gammafixed}, the transition rates for different friction coefficients $\zeta$ are plotted as the functions of the ensemble temperature $T$. It is shown that increasing $\zeta$ will slow down the transition process. This result is consistent with the kinetics of the black hole phase transition described by the Markovian models  \cite{Li:2021vdp}. In the intermediate-strong friction regime, the friction impedes the transition process.

\subsection{Oscillatory friction kernel}

In this subsection, we consider a model of the oscillatory decayed friction kernel, which is given by
\begin{eqnarray}\label{osc_friction}
\zeta(t)=\zeta e^{- |t|/\gamma}\left[\cos(\tilde{\omega}t)+\frac{1}{\tilde{\omega}\gamma}\sin(\tilde{\omega}|t|)\right]\;,
\end{eqnarray}
where $\tilde{\omega}$ is the oscillating frequency of the friction kernel, and $\gamma$ is a measure of the decay time of the memory kernel or, equivalently, of the correlation time of the fluctuations.

\begin{figure}
  \centering
  \includegraphics[width=8cm]{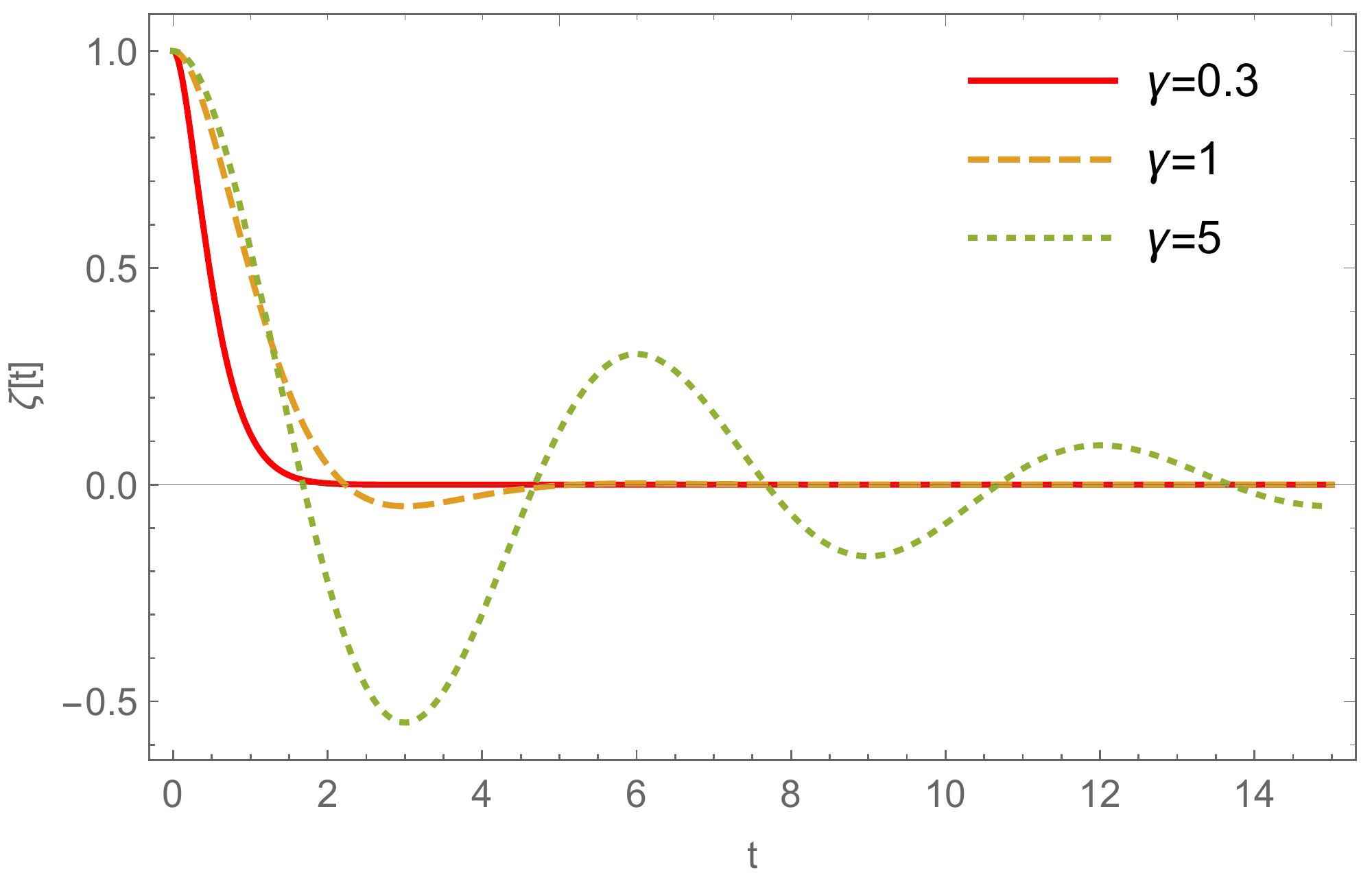}\\
  \caption{The friction kernel of the oscillating friction for different decay coefficient $\gamma$. In the plots, $\zeta=1$, and $\tilde{\omega}=\frac{\pi}{3}$.
  }\label{Oscillating_kernel}
\end{figure}

It can be seen that the oscillatory friction kernel is the exponentially decayed friction plus the oscillating factor. In Figure \ref{Oscillating_kernel}, we have plotted the oscillatory friction kernel for different decay coefficient $\gamma$. It can be seen that for small $\gamma$, the behavior is similar to the exponential decayed friction kernel. The value of $\gamma$ influences the oscillation frequency of the friction kernel but not its magnitude, which is different from the exponentially decayed friction kernel in Eq.(\ref{exp_kernel}).

In order to calculate the transition rate, we need the Laplace transform of the friction kernel (\ref{osc_friction}), which is given by  
\begin{eqnarray}\label{Osc_fric_Lap}
\hat{\zeta}(\lambda)=\frac{\zeta\gamma\left(2+\gamma\lambda\right)}{\tilde{\omega}^2\gamma^2+\left(1+\gamma\lambda\right)^2}\;.
\end{eqnarray}

\begin{figure}
  \centering
  \includegraphics[width=8cm]{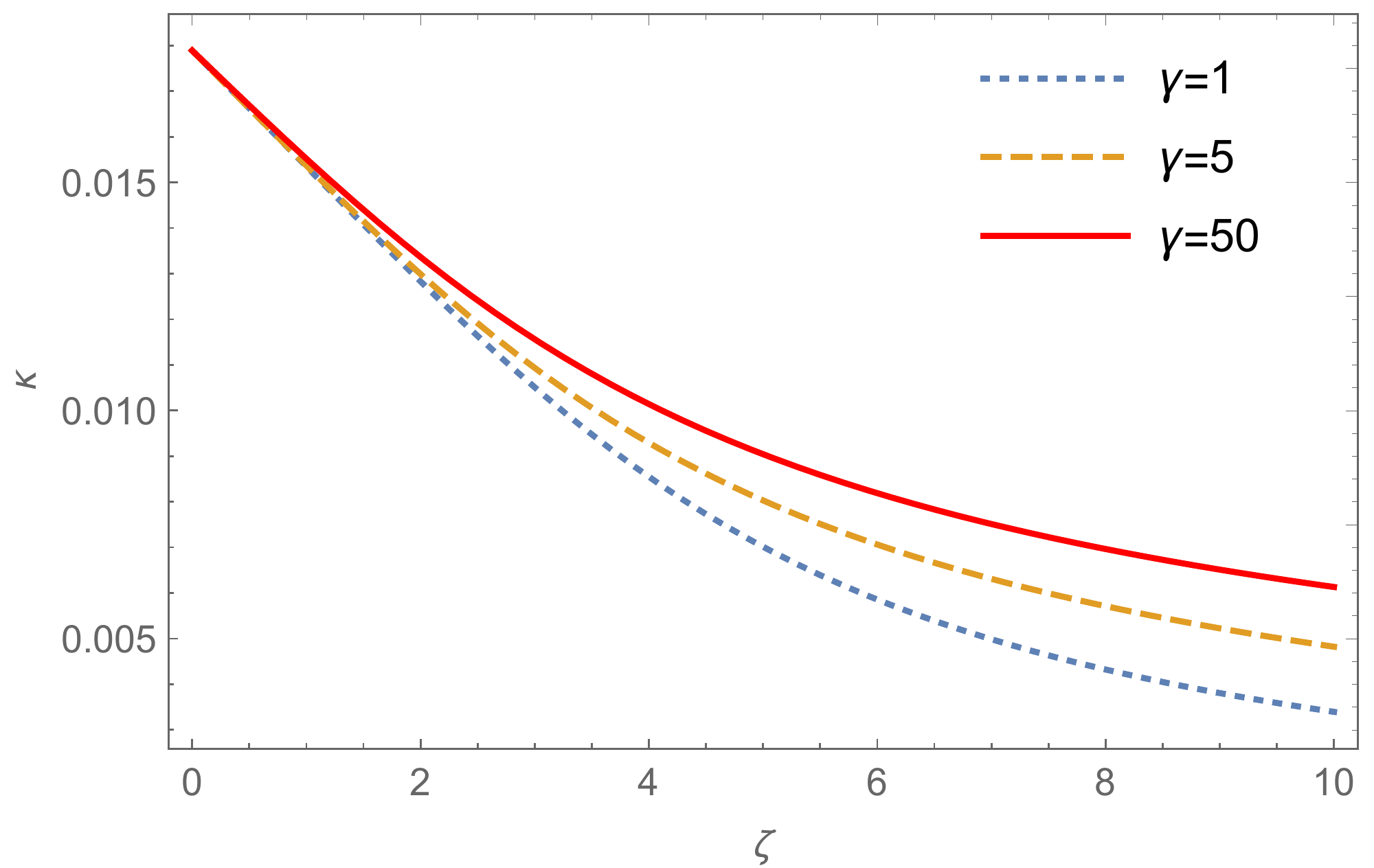}
  \includegraphics[width=8cm]{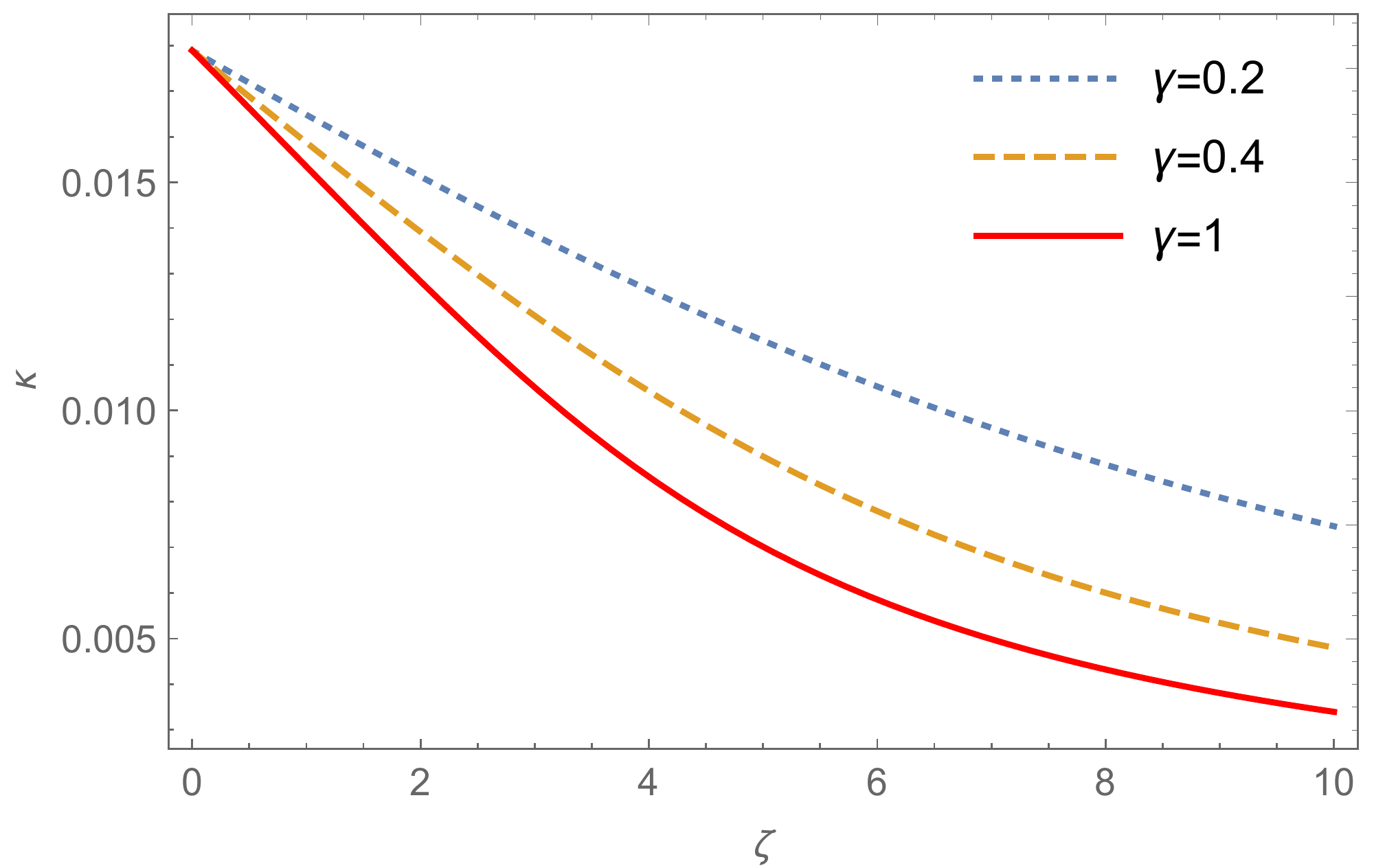}
  \caption{The dependence of the transition rate of the Hawking-Page phase transition from the large SAdS black hole state to the thermal AdS state on the friction coefficient $\zeta$ for the oscillatory friction kernel. In the plot, $L=1$, $T=0.5$, and $\tilde{\omega}=\frac{\pi}{3}$. The first panel is for $\gamma>1$ and the second panel is for $\gamma<1$.  }\label{kappa_vs_zeta_osc_HP}
\end{figure}

\begin{figure}
  \centering
  \includegraphics[width=8cm]{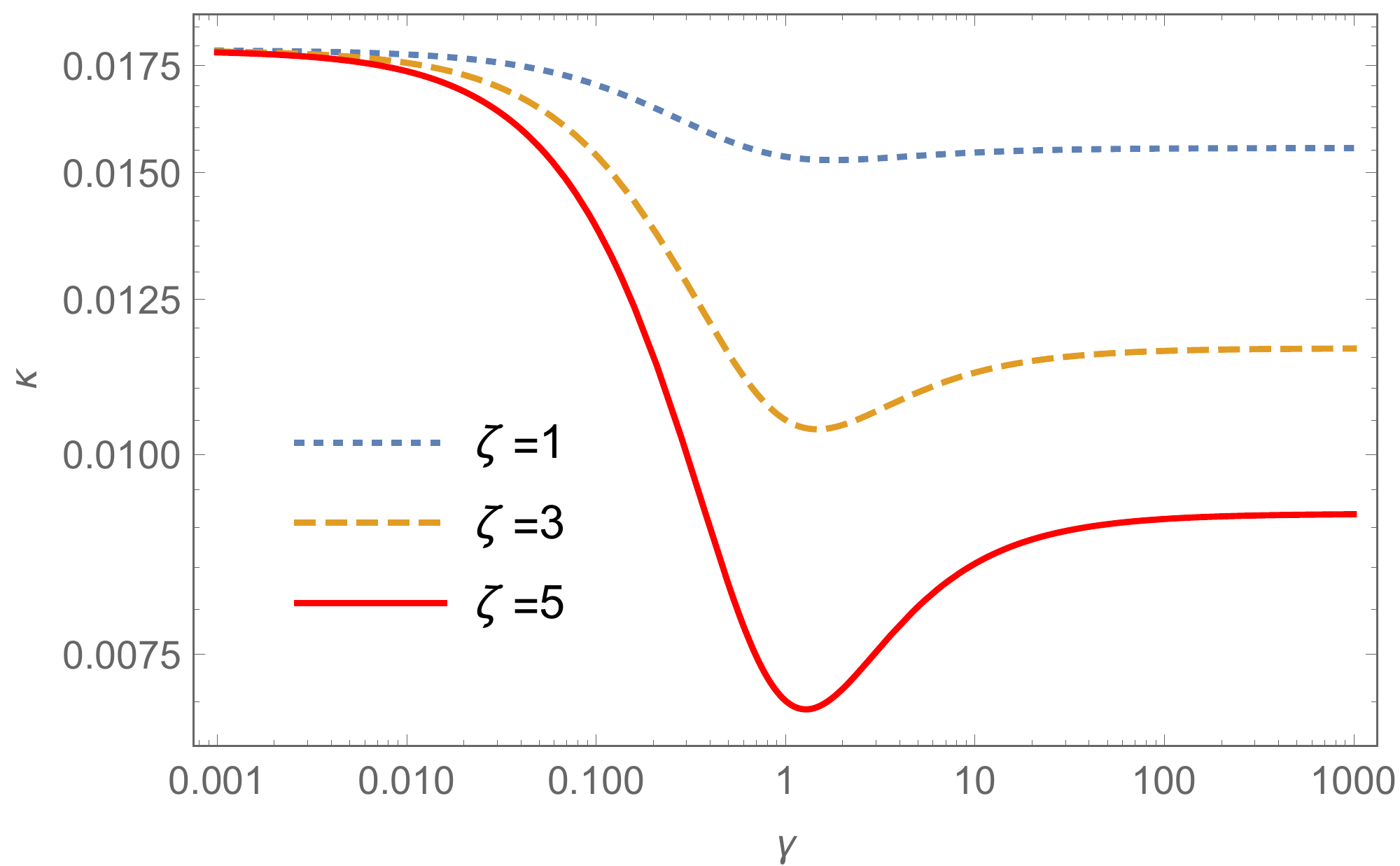}
  \caption{The dependence of the transition rate of the Hawking-Page phase transition from the large SAdS black hole state to the thermal AdS state on the decay coefficient $\gamma$ for the oscillatory friction kernel. In the plot, $L=1$, $T=0.5$, and $\tilde{\omega}=\frac{\pi}{3}$. }\label{kappa_vs_gamma_osc_HP}
\end{figure}

In Figure \ref{kappa_vs_zeta_osc_HP}, the dependence of the transition rate of the Hawking-Page phase transition from the large SAdS black hole state to the thermal AdS state on the friction coefficient $\zeta$ for the oscillatory friction kernel is plotted. It can be observed that the transition rate is a decreasing function of the friction coefficient. This means that increasing the friction slows down the transition process. However, as mentioned above, the Grote-Hynes's result of the transition rate is only valid in the intermediate-strong friction regime. We can conclude that the transition process of Hawking-Page is impeded by the friction in the intermediate-strong friction regime, which seems to be an universal property of the kinetics of the black hole phase transition not only for the Markovian model but also for the non-Markovian model. 
Another observation is that the effect of the correlation time $\gamma$ on the kinetics is not monotonic. When $\gamma>1$, increasing $\gamma$ will promote the transition process as depicted in the first panel of Figure \ref{kappa_vs_zeta_osc_HP}. When $\gamma<1$, increasing $\gamma$ will slow down the transition process as shown in the second panel of Figure \ref{kappa_vs_zeta_osc_HP}. This behavior can be explained via Figure \ref{kappa_vs_gamma_osc_HP}.

From Figure \ref{kappa_vs_gamma_osc_HP}, it can be observed that the transition rate is not the monotonic function of $\gamma$. When $\gamma<1$, the transition rate is the monotonic decreasing function of $\gamma$. When $\gamma>1$, the transition rate is the monotonic increasing function of $\gamma$. This observation is consistent with that of Figure \ref{kappa_vs_zeta_osc_HP}. There is a turnover point in the kinetics when adjusting the correlation time $\gamma$, which is different from the exponentially decayed friction kernel considered in the last subsection. The behavior of Figure \ref{kappa_vs_zeta_osc_HP} can be qualitatively explained. In the interval $\gamma<1$, increasing $\gamma$ will slow down the transition process. In the interval $\gamma>1$, increasing $\gamma$ will speed up the transition process.
Two limiting cases of $\gamma$ can be analyzed as follows. When $\gamma$ is small, expanding $\hat{\zeta}$ as the series of $\gamma$, we have
\begin{eqnarray}
\hat{\zeta}= 2\zeta \gamma+\cdots\;,
\end{eqnarray}
where only the leading term is kept. Because $\hat{\zeta}$ is irrelevant to $\lambda$, the Grote-Hynes equation  (\ref{freq_eq_HP}) becomes a quadratic equation and can be easily solved. This in turn gives the Grote-Hynes frequency $\lambda_r\simeq\omega_s-\zeta\gamma$. Then, the transition rates is given by
\begin{eqnarray}
\kappa\simeq\left(1-\frac{\zeta\gamma}{\omega_s}\right)\frac{\omega_l}{2\pi}e^{-\beta W}\;. 
\end{eqnarray}
It can be seen that when $\gamma$ is small, the transition rate is the decreasing function of $\gamma$. When $\gamma\rightarrow0$, the transition rate for different $\zeta$ have an universal value, which is given by $\kappa=\frac{\omega_l}{2\pi}e^{-\beta W}$. The analytical results are consistent with the plots in Figure \ref{kappa_vs_gamma_osc_HP}. 

When $\gamma$ is large, we have 
\begin{eqnarray}
\hat{\zeta}(\lambda)=\frac{\zeta\lambda}{\tilde{\omega}^2+\lambda^2}+\frac{2\zeta \tilde{\omega}^2}{(\tilde{\omega}^2+\lambda^2)^2}\frac{1}{\gamma}+\cdots\;.
\end{eqnarray} 
In this case, the Grote-Hynes equation (\ref{freq_eq_HP}) is a higher-degree polynomial equation and the analytical solution of the Grote-Hynes frequency is impossible. In the limit $\gamma\rightarrow \infty$, $\hat{\zeta}(\lambda)$ is relevant to the two parameters $\zeta$ and $\tilde{\omega}$, and is irrelevant to $\gamma$. Then, the Grote-Hynes frequency is also determined by $\zeta$ and $\tilde{\omega}$. Therefore, in the limit $\gamma\rightarrow \infty$, the transition rates for different friction coefficients $\zeta$ have different limiting values, as shown in Figure \ref{kappa_vs_gamma_osc_HP}. In summary, in the limiting case of $\gamma\rightarrow 0$, the non-Markovian effects are eliminated, while in the limit of $\gamma\rightarrow \infty$, the non-Markovian effects still influence the kinetics of Hawking-Page phase transition.    

In fact, the tendency of the plots in Figure \ref{kappa_vs_gamma_osc_HP} can also be analyzed directly from the derivative of $\lambda$ by regarding $\lambda$ as the function of $\gamma$. The derivative can be calculated by combining the the Grote-Hynes equation (\ref{freq_eq_HP}) and the Laplace transformation of the oscillating kernel Eq.(\ref{Osc_fric_Lap}). The result is given by  
\begin{eqnarray}
\frac{d\lambda}{d\gamma} \propto \zeta(-1-\lambda\gamma+\hat{\omega}^2\gamma^2)\;,
\end{eqnarray}
where the omitted cumbersome prefactor is positive. When $\gamma$ is small enough, the dominant contribution to the derivative is the first term on the right hand and the derivative is negative. The transition rate, which is proportional to $\lambda$, is therefore the decreasing function of $\gamma$. This is consistent with our previous analytical results. When $\gamma$ is large enough, the dominant term of the derivative is the third term on the right hand and the transition rate is the increasing function of $\gamma$ due to the positive derivative. This analysis gives the mathematical interpretation of the non-monotonic behavior for the transition rate in Figure \ref{kappa_vs_gamma_osc_HP}.

\begin{figure}
  \centering
  \includegraphics[width=8cm]{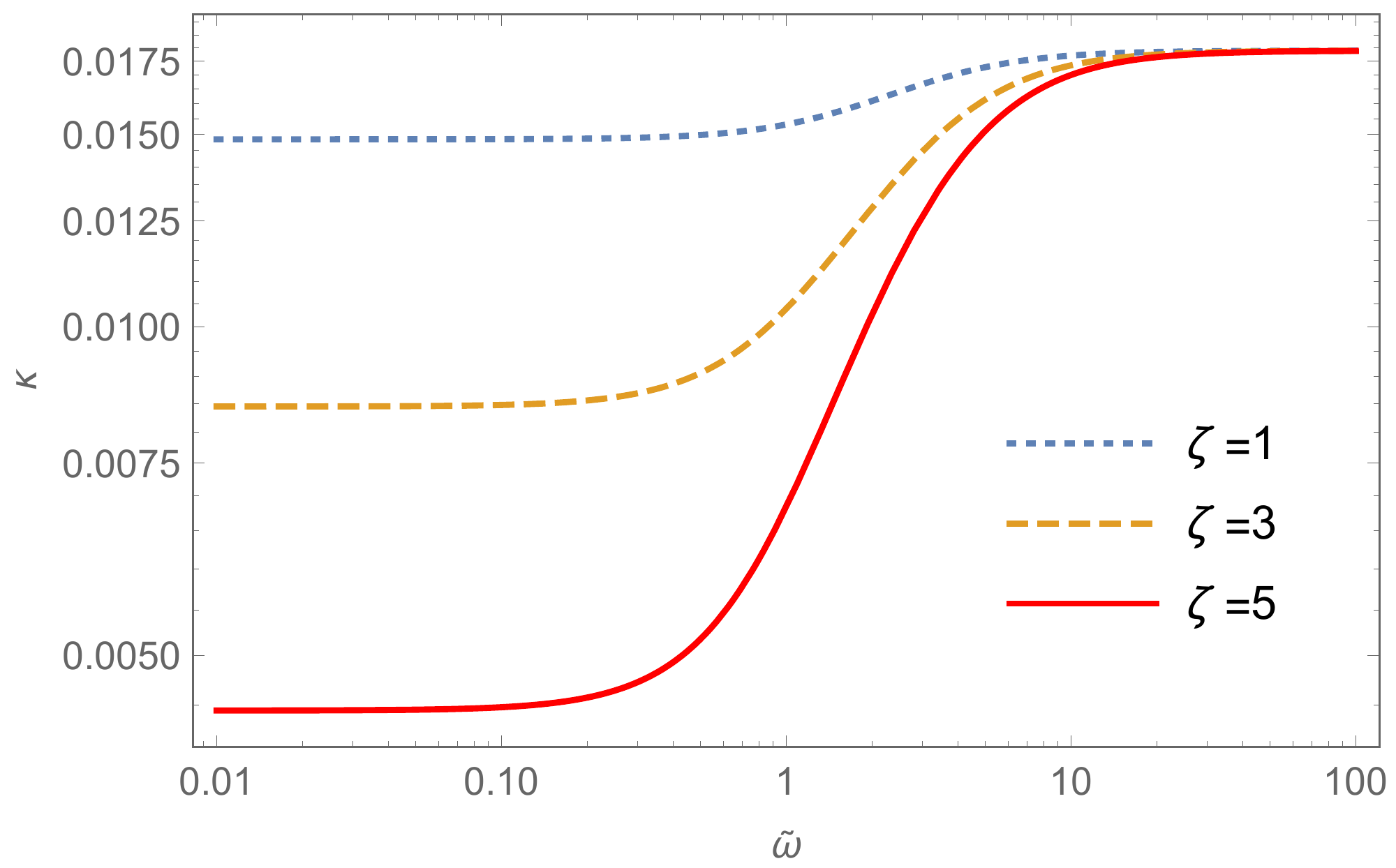}
  \caption{The dependence of the transition rate of the Hawking-Page phase transition from the large SAdS black hole state to the thermal AdS state on the oscllating frequency $\tilde{\omega}$ for the oscillatory friction kernel. In the plot, $L=1$, $T=0.5$, and $\gamma=1$. }\label{kappa_vs_omega_osc_HP}
\end{figure}

The dependence of the transition rate of the Hawking-Page phase transition from the large SAdS black hole state to the thermal AdS state on the oscillating frequency $\tilde{\omega}$ for the oscillatory friction kernel is plotted in Figure \ref{kappa_vs_omega_osc_HP}. It can be observed that the transition rate is a monotonic function of the oscillating frequency. Increasing $\tilde{\omega}$ will speed up the transition process. One can also analyze the two limiting cases where $\tilde{\omega}\rightarrow 0$ and $\tilde{\omega}\rightarrow \infty$. When $\tilde{\omega}\rightarrow 0$, $\hat{\zeta}(\lambda)\rightarrow \frac{\zeta\gamma(2+\gamma\lambda)}{(1+\gamma\lambda^2)}$, which is irrelevant to $\tilde{\omega}$. Therefore, the transition rates for different $\zeta$ have different limiting values. When $\tilde{\omega}\rightarrow \infty$, $\hat{\zeta}(\lambda)\rightarrow 0$. The transition rate has an universal limiting value, which is given by $\kappa=\frac{\omega_l}{2\pi}e^{-\beta W}$. One can conclude that the non-Markovian effects are eliminated in the limit of $\tilde{\omega}\rightarrow \infty$ and survived in the limit of $\tilde{\omega}\rightarrow 0$.

\begin{figure}
  \centering
  \includegraphics[width=8cm]{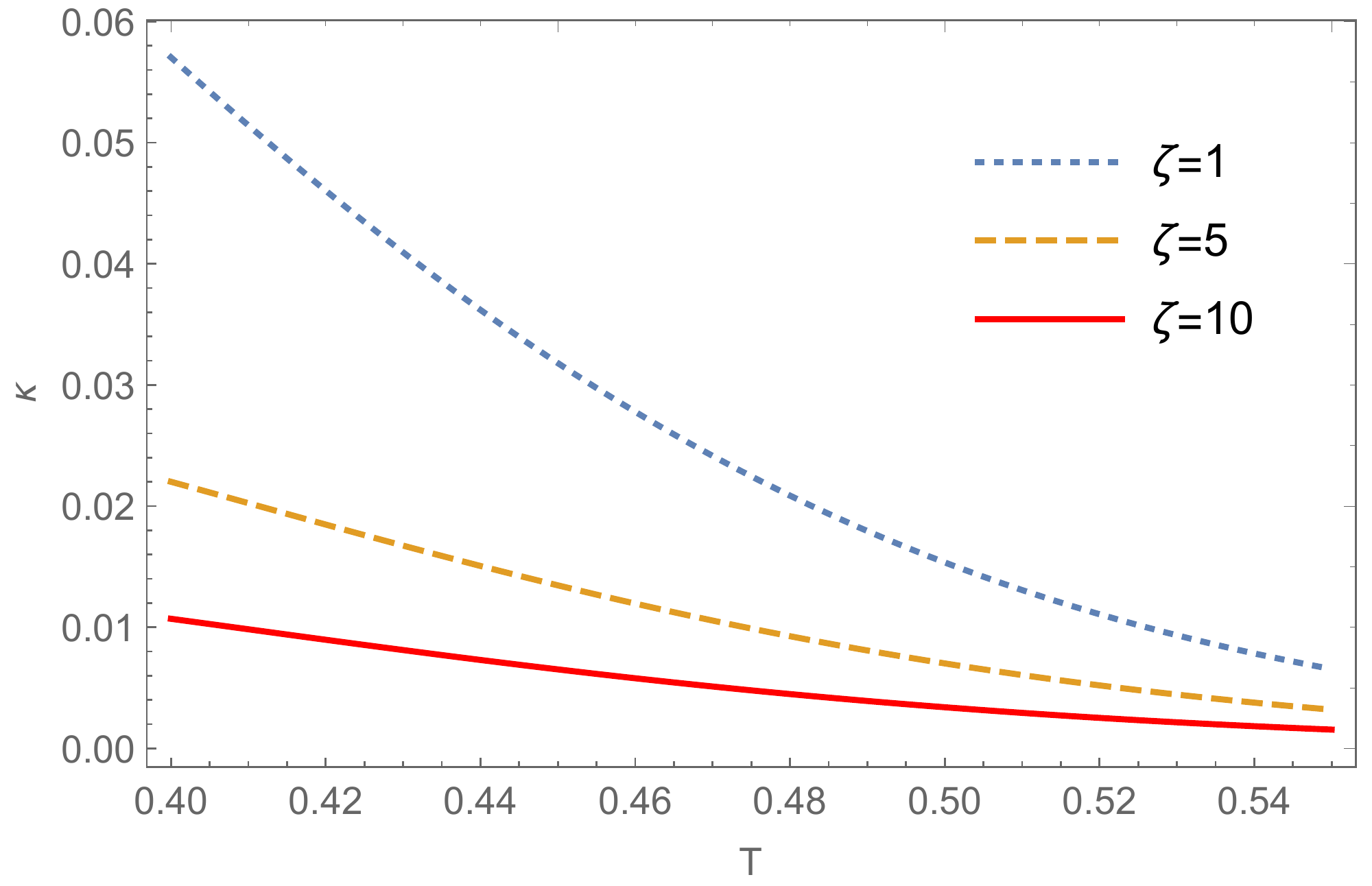}\\
  \caption{The dependence of the transition rate of the Hawking-Page phase transition from the large SAdS black hole state to the thermal AdS state on the ensemble temperature $T$ for the exponentially decayed friction kernel. In the plot, $L=1$, $\gamma=1$, and $\tilde{\omega}=\frac{\pi}{3}$. }\label{kappa_vs_T_osc_HP_diffzeta}
\end{figure}

\begin{figure}
  \centering
  \includegraphics[width=8cm]{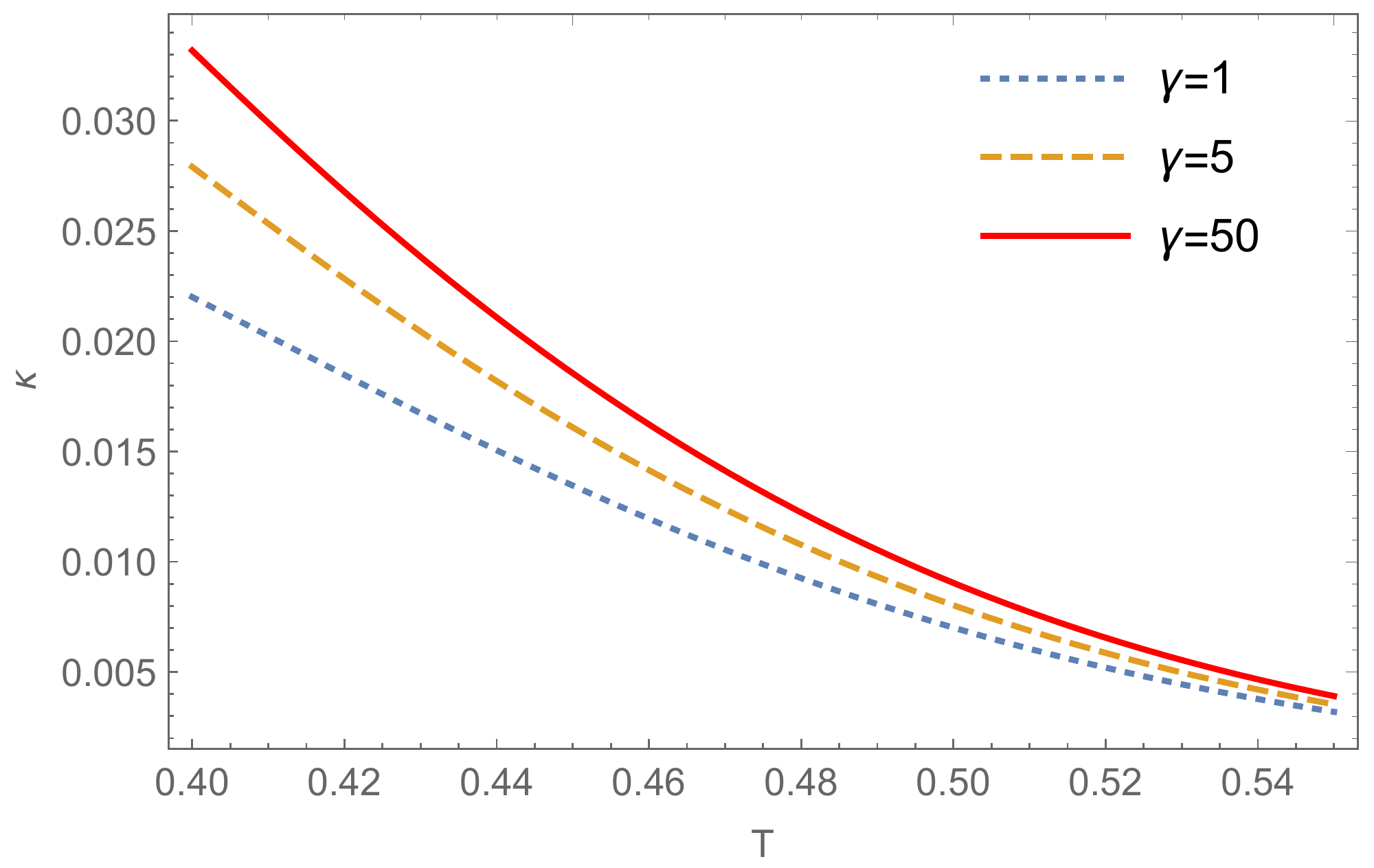}
  \includegraphics[width=8cm]{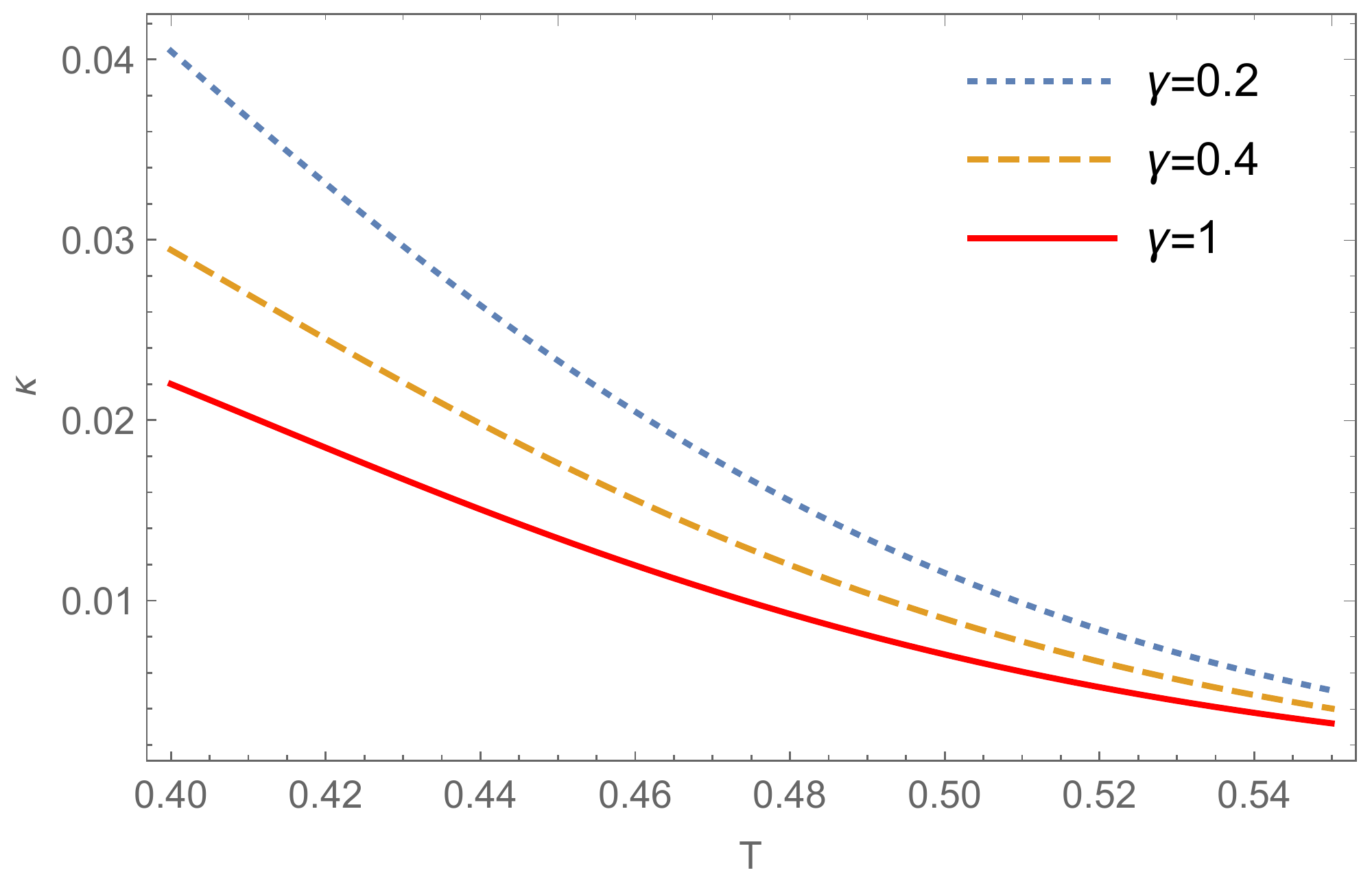}\\
  \caption{The dependence of the transition rate of the Hawking-Page phase transition from the large SAdS black hole state to the thermal AdS state on the ensemble temperature $T$ for the exponentially decayed friction kernel. In the plot, $L=1$,  $\zeta=5$, and $\tilde{\omega}=\frac{\pi}{3}$. }\label{kappa_vs_T_osc_HP_diffgamma}
\end{figure}

\begin{figure}
  \centering
  \includegraphics[width=8cm]{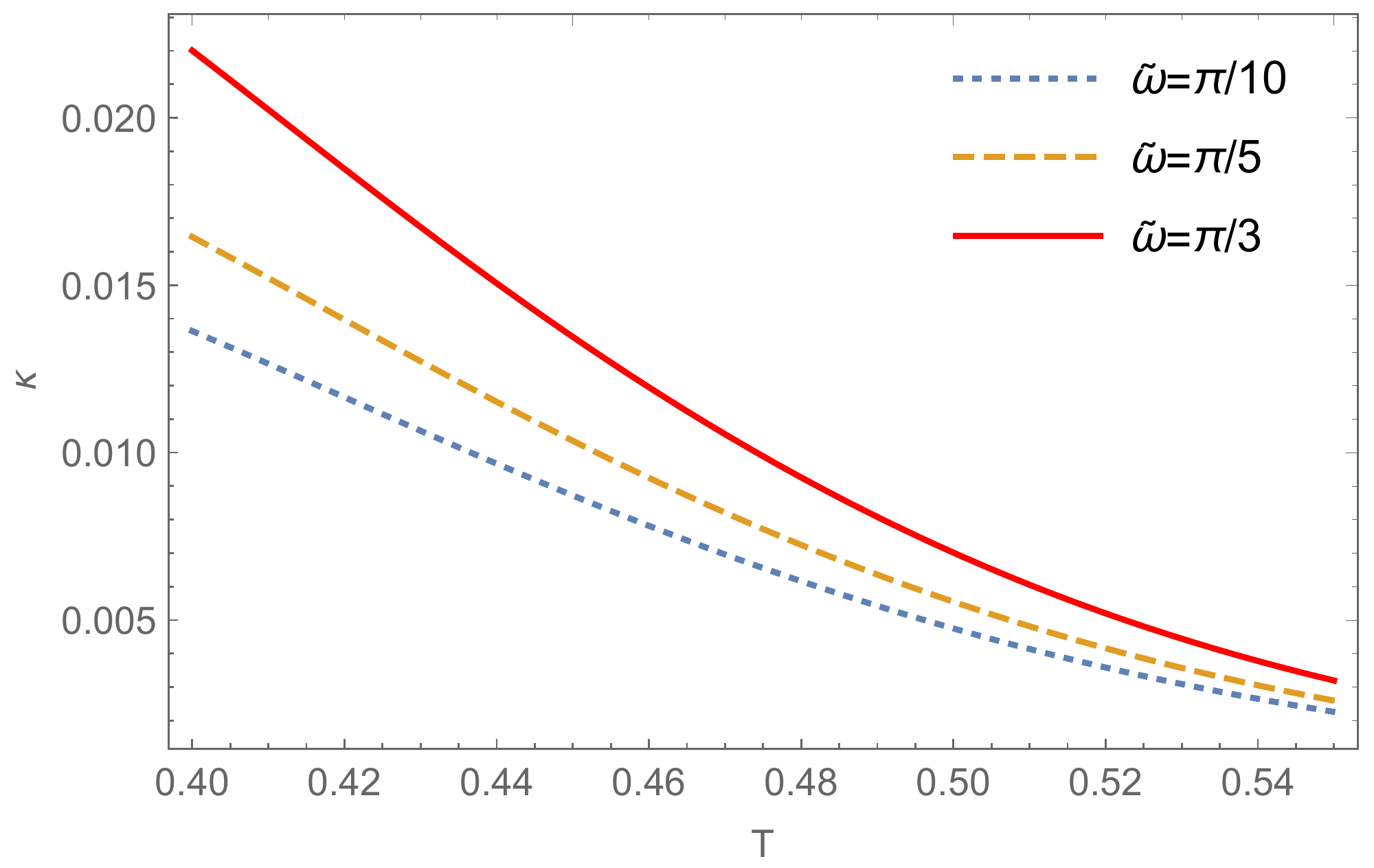}\\
  \caption{The dependence of the transition rate of the Hawking-Page phase transition from the large SAdS black hole state to the thermal AdS state on the ensemble temperature $T$ for the exponentially decayed friction kernel. In the plot, $L=1$,  $\zeta=5$, and $\gamma=1$. }\label{kappa_vs_T_osc_HP_diffomega}
\end{figure}

In Figure \ref{kappa_vs_T_osc_HP_diffzeta}-\ref{kappa_vs_T_osc_HP_diffomega}, we plot the transition rates as the function of the ensemble temperature. In Figure \ref{kappa_vs_T_osc_HP_diffzeta}, the correlation time $\gamma$ and the oscillating frequency $\tilde{\omega}$ are kept fixed, and the friction strength $\zeta$ are varied. In Figure \ref{kappa_vs_T_osc_HP_diffgamma}, the friction strength  and the oscillating frequency $\tilde{\omega}$ are kept fixed, and the correlation time $\gamma$ are varied. In Figure \ref{kappa_vs_T_osc_HP_diffomega}, the correlation time $\gamma$ and the friction strength are kept fixed, and the oscillating frequency $\tilde{\omega}$ are varied.  
The first observation from these plots is that the transition rate is a decreasing function of the ensemble temperature. The reason is that the kinetics is exponentially related to the barrier height between the small SAdS black hole state at the top of the barrier and the large SAdS black hole state in the right potential well on the landscape. Increasing the ensemble temperature will elevate the barrier height and in turn slow down the dynamical process of Hawking-Page phase transition. From Figure \ref{kappa_vs_T_osc_HP_diffzeta}, we observe that increasing the friction strength $\zeta$ will slow down the kinetics. From Figure \ref{kappa_vs_T_osc_HP_diffgamma}, we find that the kinetics affected by the correlation time $\gamma$ of the effective thermal bath in different regime shows different behavior. For $\gamma>1$ (the first panel of \ref{kappa_vs_T_osc_HP_diffgamma}), increasing the correlation time speeds up the transition process, while for $\gamma<1$ (the second panel of \ref{kappa_vs_T_osc_HP_diffgamma}), increasing the correlation time slows down the transition process instead. In Figure \ref{kappa_vs_T_osc_HP_diffomega}, one can conclude that the larger oscillating frequency makes the transition process easier. All these observations are consistent with the previous plots.

\section{Conclusion and discussion}
\label{Con_dis}

In conclusion, based on the free energy landscape description of Hawking-Page phase transition, we have studied the non-Markovian effects on the kinetics of Hawking-Page phase transition. As a simple application of the kinetic result, we focus on the transition process from the large SAdS black hole state to the thermal AdS space state on the free energy landscape. From the analytical and numerical results, there are certain universal properties of the kinetics of Hawking-Page phase transition. The first is that kinetics represented by the transition rate is exponentially related to the barrier height and in turn depends on the ensemble temperature because the free energy landscape of hawking-Page phase transition is modulated globally by the ensemble temperature. The second one is that increasing the friction strength will slow down the transition process. This appears to be valid in the intermediate-strong friction regime not only for the Markovian dynamics but also for the non-Markovian dynamics. For the exponentially decayed friction kernel, the non-Markovian effects promote the transition process, and for the oscillatory decayed friction kernel, increasing the oscillating frequency can also speed up the transition process.

At last, we discuss three possible aspects that deserve further investigation. In the present work, we only discuss Hawking-Page phase transition. One can generalize the current work to study other types of phase transition processes in black hole physics. The analytical method is employed to study the non-Markovian dynamics of Hawking-Page phase transition. Note that the generalized Langevin equation is an integro-differential equation. Solving the generalized Langevin equation numerically will give the exact results of the kinetics of the black hole phase transition, which is still challenging at the moment. The last one is that the kinetic theory considered in the present work is only valid in the intermediate-strong friction regime. For the Markovian model, it has been shown that there is a kinetic turnover point when varying the friction strength \cite{Li:2021vdp}. It is quite interesting to consider the non-Markovian effects on the kinetics of black hole phase transition in the weak friction regime.

\end{document}